\documentclass[preprintnumbers]{revtex4}
\UseRawInputEncoding
\usepackage{amssymb}
\usepackage{amsmath}
\usepackage{graphicx}
\usepackage{calligra}
\usepackage{mathrsfs}
\usepackage{dcolumn}
\usepackage{bm}
\usepackage{subfigure}
\usepackage{color}
\usepackage{CJKutf8}
\usepackage{float}
\numberwithin{equation}{section}
\numberwithin{figure}{section}

\begin{document}
\title{Schottky Anomaly of five-Dimensional de Sitter Hairy Spacetime based on effective thermodynamic quantities}
\author{Shan-Xia Bao$^{1,2}$, Ren Zhao$^{1}$, Yun-Zhi Du$^{1,2,3}$}
\thanks{\emph{e-mail:duyzh22@sxdtdx.edu.cn }(corresponding author)}

\address{$^1$Department of Physics, Shanxi Datong University, Datong 037009, China\\
$^2$Institute of Theoretical Physics, Shanxi Datong University, Datong, 037009, China\\
$^3$State Key Laboratory of Quantum Optics and Quantum Optics Devices, Shanxi University, Taiyuan 030006, China\\}

\begin{abstract}

Taking the mass, charge, hair parameter, and cosmological constant of the 5-dimensional de Sitter hairy spacetime as the state variables of a thermodynamic system, and based on the satisfaction of the universal first law of thermodynamics, we obtain the effective thermodynamic quantities of the spacetime. The thermodynamic properties of the system in the coexistence region of the black hole and cosmological horizons are discussed. We find that under certain conditions, the heat capacity of the effective thermodynamic system in the two-horizon coexistence region of de Sitter hairy spacetime, as a function of either temperature or the ratio of the horizon positions, exhibits a peak-like behavior similar to that observed in paramagnetic systems. Further analysis reveals that, under specific conditions, the heat capacity of this effective thermodynamic system in dS spacetime resembles that of a two-level system composed of two horizons with different radiation temperatures. By comparing these results, we derive the expressions for the number of microscopic particles on the two horizons within the coexistence region. This outcome reflects the quantum nature of dS spacetime and provides a new pathway for further in-depth investigation into the thermodynamic and quantum properties of the two-horizon coexistence region in dS spacetime.

\end{abstract}

\maketitle

\section{Introduction}\label{one}

The proposal of black hole thermodynamics has, to some extent, served as a bridge and link connecting general relativity, classical thermodynamics, and quantum mechanics. With the establishment of the four laws of black hole thermodynamics, the studies of black hole thermodynamics and its thermal properties have become one of the most active topics in theoretical physics research \cite{1,2,3,4,5,6,7,8,9,10,11,12,13,14,15}.

It is well known that phase transitions in ordinary thermodynamic systems result from the competition among their microscopic constituents. Since black holes exhibit thermal behaviors analogous to those of ordinary thermodynamic systems, investigating their internal microstructure has become a subject of significant interest. George Ruppeiner investigated the microstructure of AdS black holes using the Riemannian curvature scalar within the framework of Ruppeiner geometry \cite{16}. Through introducing the number density of black hole molecules to examine their phase transitions and microstructure, he demonstrated that different values of the Riemannian curvature scalar correspond to different types of interactions among black hole molecules \cite{17}. This approach was promptly extended to studies of other black holes \cite{18,19,20}. Other researchers have also explored the internal microstructure of black holes using various methods, achieving considerable progress \cite{21,22,23,24,25,26,27}. Recently, Ref. \cite{28} proposed employing non-extensive R\'{e}nyi statistics to investigate the microstructure of asymptotically flat charged black holes. Ref. \cite{29} utilized Ruppeiner geometry to look into the AdS black hole's microscopic structure and numerically computed the Ricci curvature scalar $R$ to explain the interactions between the AdS black hole's microscopic particles under the influence of $f(Q)$ gravity. Ref. \cite{30} investigated the microstructures of a charged AdS black hole exhibiting quantum anomalies through the lens of Ruppeiner geometry. Ref. \cite{31} studied the microstructure of the FRW universe as a thermodynamic system, providing new insights into the interactions among the ideal fluid components within the expanding FRW universe.

In recent years, with in-depth studies of dark energy and dark matter, the thermodynamic properties of de Sitter spacetime have attracted significant attention \cite{32,33,34,35,36,37,38,39,40,41,42}. Latest astronomical observations indicate that approximately 70\% of the energy density in the universe consists of so-called dark energy. In general relativity, this dark energy is described by the cosmological term originally introduced by Einstein. Analysis of supernova data further suggests the existence of a positive cosmological constant in the universe. During the early inflationary period, our universe exhibited characteristics of a quasi-de Sitter spacetime. The cosmological constant introduced in de Sitter spacetime studies originates from vacuum energy contributions, which itself represents a form of material energy. If the cosmological constant indeed corresponds to dark energy, our universe would eventually evolve into a new de Sitter phase \cite{43}. The observed accelerated expansion of the universe also indicates its asymptotic approach to a de Sitter spacetime. Furthermore, the success of AdS/CFT correspondence has motivated searches for similar dual relationships in de Sitter space. Therefore, investigating de Sitter spacetime holds both theoretical significance and practical necessity.

As is well known, in de Sitter spacetime, when the state parameters of the spacetime satisfy certain conditions, both a black hole event horizon and a cosmological event horizon exist. Each horizon possesses its corresponding Hawking radiation temperature, which are generally not equal. However, since both the radiation temperatures and the thermodynamic quantities associated with the two horizons are functions of the spacetime parameters, these thermodynamic quantities are not independent but exhibit certain correlations. A significant question in contemporary theoretical physics is whether a single thermodynamic system can effectively describe the thermodynamic properties of the region where both horizons coexist. Considerable work has been conducted in this area in recent years \cite{36,37,40,41,42,43}, yielding notable results. Under current understanding, any physical system must obey the universal first law of thermodynamics. This principle implies that a thermodynamic system constructed to reflect the properties of the two-horizon coexistence region in de Sitter spacetime must consistently satisfy the first law. Guided by this idea, a differential equation governing the effective temperature of the thermodynamic system for the two-horizon coexistence region has been derived. By applying appropriate boundary conditions for this region and solving the equation, the effective thermodynamic quantities for a spherically symmetric charged spacetime have been obtained \cite{44,45}. Building upon this foundation, this paper investigates the effective thermodynamic quantities and the corresponding thermodynamic properties described by these quantities for the universally applicable 5-dimensional de Sitter hairy spacetime.

In this study, we observe that within the 5-dimensional de Sitter hairy spacetime, the heat capacity of the effective thermodynamic system in the two-horizon coexistence region exhibits a Schottky-like anomaly in both the canonical and grand canonical ensembles. The variation of heat capacity with temperature resembles that of an ordinary two-level thermodynamic system. This suggests that the two horizons in the coexistence region-each with a distinct radiation temperature-can be regarded as two different energy levels in an effective thermodynamic system. When $N$ particles reside between the two horizons, they form a two-level system. A comparison of the heat capacity curves among the effective thermodynamic system, the ordinary two-level thermodynamic system, and the two-horizon system reveals a remarkable similarity. By comparing the temperature-dependent heat capacity curve of the effective system in the two-horizon coexistence region with that of a two-level system treating the two horizons as distinct energy levels, we can deduce the number of microscopic particles between the horizons. This finding not only deepens our understanding of de Sitter spacetime but also provides a new approach to study microscopic interactions within black holes and simulate cosmic evolution.

The structure of this paper is as follows: In Section 2, we present the thermodynamic quantities corresponding to the black hole horizon and the cosmological horizon in a 5-dimensional de Sitter hairy spacetime. In Section 3, under given boundary conditions, we discuss the effective thermodynamic quantities in the coexisting region of the black hole horizon and the cosmological horizon in this spacetime. In Section 4, we analyze the heat capacities of the canonical and grand canonical ensembles-where the electric charge or electric potential is held constant, respectively-in the two-horizon coexisting region. The influence of different spacetime parameters on the heat capacity is also examined. In Section 5, we first review the heat capacity curve of an ordinary paramagnetic system, i.e., a two-level system. By comparing the heat capacity curve of the thermodynamic system in the two-horizon coexisting region with that of the two-level system, we derive a relation between the number of microscopic particles on the horizons and the state parameters. Finally, in Section 6, we provide a summary and discussion.(We use the units $G=\hbar ={{k}_{B}}=c=1$).

\section{5-Dimensional de Sitter Hairy Spacetime} \label{two}

Although our primary interest lies in Einstein gravity, we frame our investigation within the broader context of Lovelock gravity minimally coupled to a Maxwell field. The scalar hair is conformally coupled to gravity through the dimensionally extended Euler densities, expressed in terms of a rank-four tensor \cite{46}. The spherically symmetric black hole solution in a topological spacetime is given by:
\begin{align}\label{2.1}
 d{{s}^{2}}=-f(r)d{{t}^{2}}+{{f}^{-1}}d{{r}^{2}}+{{r}^{2}}d\Omega _{2}^{2}
\end{align}
with the horizon function:
\begin{align}\label{2.2}
f(r)=1-\frac{8M}{3\pi {{r}^{2}}}-\frac{\Lambda {{r}^{2}}}{6}-\frac{H}{{{r}^{3}}}+\frac{4\pi {{Q}^{2}}}{3{{r}^{4}}}
\end{align}
where $M$, $H$ and $Q$ are spacetime mass, hair parameter and charge respectively, $\Lambda$  is the cosmological constant.

From the condition $f(r)=0$, the relationship between the black hole mass $M$ and the horizon radius $r$ can be derived as:
\begin{align}\label{2.3}
M=\frac{3\pi {{r}^{2}}}{8}-\frac{3\pi \Lambda {{r}^{4}}}{48}-\frac{3\pi H}{8r}+\frac{{{\pi }^{2}}{{Q}^{2}}}{2{{r}^{2}}}
\end{align}
\begin{figure}[htb]
\centering
\includegraphics[width=8cm,height=4.5cm]{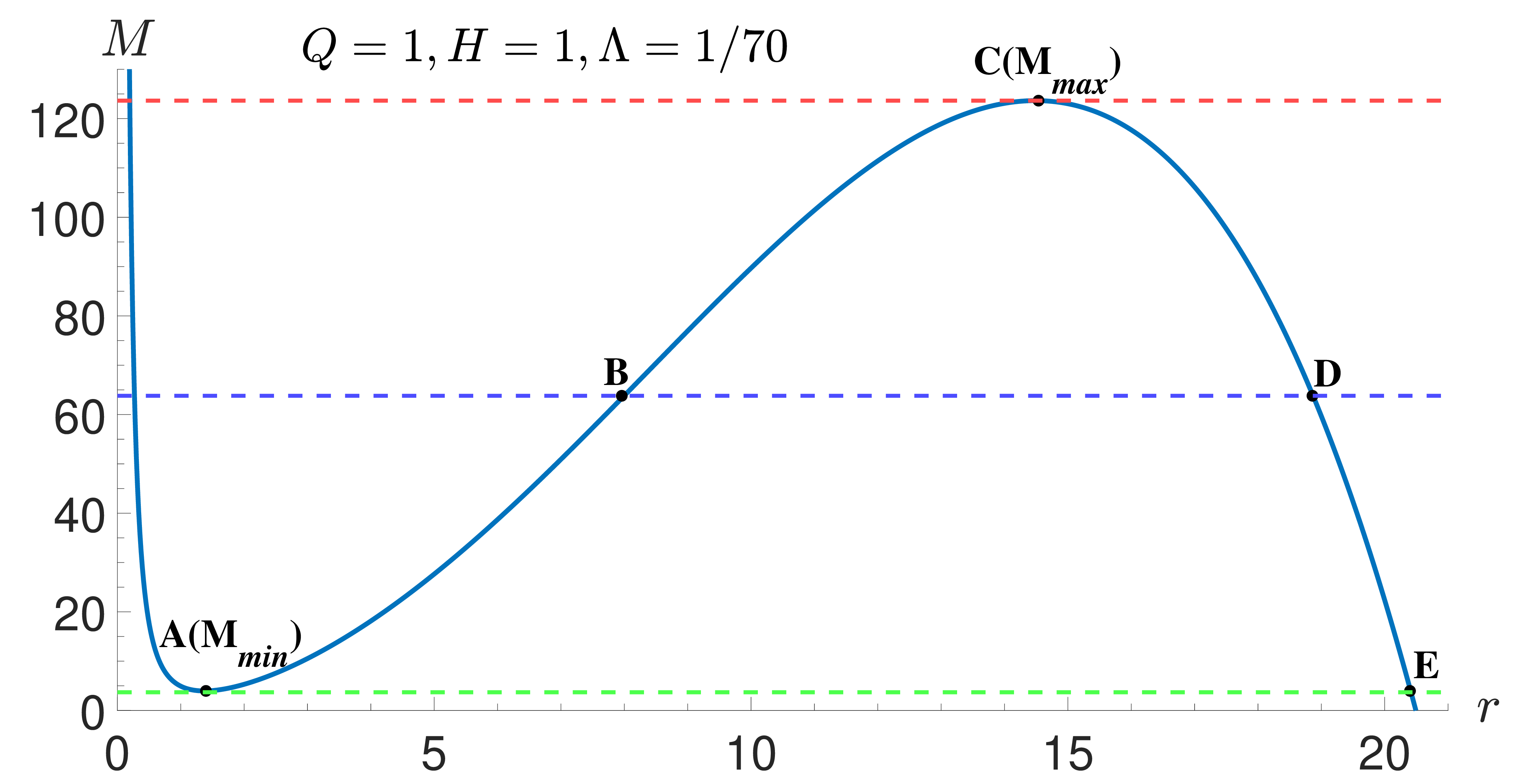}
\caption{(Color online) The curve for fixed parameters $Q$=1, $H$=1 and $\Lambda$=1/70}
\end{figure}

Figure 2.1 reveals a common characteristic in the curve of black hole mass versus horizon radius:
\begin{align}\label{2.4}
M(r\to 0)\to \infty ~~ and ~~ M(r\to \infty )\to -\infty
\end{align}

In addition, for the appropriate values of the charge $Q$, hair parameter $H$ and the cosmological constant $\Lambda$, the black hole mass curve displays one local minimum following one local maximum, both residing above the horizontal axis. This configuration indicates the possible presence of three horizons: the inner horizon ${{r}_{-}}$, the event horizon ${{r}_{+}}$ (${{r}_{+}}\ge {{r}_{-}}$), and the cosmological horizon ${{r}_{c}}$ (${{r}_{c}}\ge {{r}_{+}}$). When the local extremum points merge into an inflection point, these three horizons coincide. In this case, known as an ultracold black hole, the horizon radius ${{r}_{ucold}}$, for given charge $Q$ and hair parameter $H$, is given by:
\begin{align}\label{2.5}
12r_{ucold}^{4}+15H{{r}_{ucold}}=40\pi {{Q}^{2}}
\end{align}

For given charge $Q$, hair parameter $H$, and cosmological constant $\Lambda$, the mass $M$ of a spacetime possessing both a black hole horizon ${{r}_{+}}$ and a cosmological horizon ${{r}_{c}}$ falls within the range $M\in [{{M}_{\min }},{{M}_{\max }}]$. As $M={{M}_{\max }}$, the two horizons ${{r}_{+}}$ and ${{r}_{c}}$ coincide together. This configuration is known as the Nariai black hole, and its event horizon radius ${{r}_{N}}$ is given by the largest positive real solution of the following equation \cite{47}:
\begin{align}\label{2.6}
6{{r}^{4}}-2\Lambda {{r}^{6}}+3Hr=8\pi {{Q}^{2}}
\end{align}

As $M={{M}_{\min}}$ the inner horizon ${{r}_{-}}$ and the black hole horizon ${{r}_{+}}$ coincide together, and such a black hole is called the cold black hole whose event horizon radius ${{r}_{cold}}$ is the remaining positive real solution of Eq. (2.6).

Then from Eq. (2.3),we can obtain:
\begin{align}\label{2.7}
M=\frac{3\pi r_{+}^{2}}{8}-\frac{3\pi \Lambda r_{+}^{4}}{48}-\frac{3\pi H}{8{{r}_{+}}}+\frac{{{\pi }^{2}}{{Q}^{2}}}{2r_{+}^{2}}
\end{align}

\begin{align}\label{2.8}
M=\frac{3\pi r_{c}^{2}}{8}-\frac{3\pi \Lambda r_{c}^{4}}{48}-\frac{3\pi H}{8{{r}_{c}}}+\frac{{{\pi }^{2}}{{Q}^{2}}}{2r_{c}^{2}}
\end{align}

The Hawking radiation temperatures for the black hole event horizon and the cosmological horizon, derived from Eq. (2.2), are respectively given by:

\begin{align}\label{2.9}
{{T}_{+}}=\frac{1}{2\pi {{r}_{+}}}-\frac{\Lambda {{r}_{+}}}{6\pi }-\frac{2{{Q}^{2}}}{3r_{+}^{5}}+\frac{H}{4\pi r_{+}^{4}}
\end{align}
\begin{align}\label{2.10}
{{T}_{c}}=-\frac{1}{2\pi {{r}_{c}}}+\frac{\Lambda {{r}_{c}}}{6\pi }+\frac{2{{Q}^{2}}}{3r_{c}^{5}}-\frac{H}{4\pi r_{c}^{4}}
\end{align}
here $\Lambda =-8\pi P$. The thermodynamic quantities associated with the black hole horizon satisfy both the first law of thermodynamics and the Smarr relation \cite{46}:

\begin{align}\label{2.11}
\delta M={{T}_{+}}\delta {{S}_{+}}+{{V}_{+}}\delta P+{{\phi }_{+}}\delta Q+{{k}_{+}}\delta H
\end{align}
\begin{align}\label{2.12}
2M=3{{T}_{+}}{{S}_{+}}-2{{V}_{+}}P+2{{\phi }_{+}}Q+3{{k}_{\text{+}}}H
\end{align}
where, the corresponding entropy ${{S}_{+}}$, thermodynamic volume ${{V}_{+}}$, electric potential ${{\phi }_{+}}$ and ${{k}_{+}}$ of the black hole horizon are:
\begin{align}\label{2.13}
{{S}_{+}}=\frac{{{\pi }^{2}}r_{+}^{3}}{2}, {{V}_{+}}=\frac{{{\pi }^{2}}r_{+}^{4}}{2}, {{\phi }_{+}}=\frac{{{\pi }^{2}}Q}{r_{+}^{2}}, {{k}_{+}}=-\frac{3\pi }{8{{r}_{+}}}
\end{align}

The thermodynamic quantities associated with the cosmological horizon satisfy both the first law of thermodynamics and the Smarr relation \cite{46}:
\begin{align}\label{2.14}
\delta M=-{{T}_{c}}\delta {{S}_{c}}+{{V}_{c}}\delta P+{{\phi }_{c}}\delta Q+{{k}_{c}}\delta H
\end{align}
\begin{align}\label{2.15}
2M=-3{{T}_{c}}{{S}_{c}}-2{{V}_{c}}P+2{{\phi }_{c}}Q+3{{k}_{c}}H
\end{align}
where, the corresponding entropy ${{S}_{c}}$, thermodynamic volume ${{V}_{c}}$, electric potential ${{\phi }_{c}}$ and ${{k}_{c}}$ of the cosmological horizon are:
\begin{align}\label{2.16}
{{S}_{c}}=\frac{{{\pi }^{2}}r_{c}^{3}}{2}, {{V}_{c}}=\frac{{{\pi }^{2}}r_{c}^{4}}{2}, {{\phi }_{c}}=\frac{{{\pi }^{2}}Q}{r_{c}^{2}}, {{k}_{c}}=-\frac{3\pi }{8{{r}_{c}}}
\end{align}

Then, from Eqs. (2.7) and (2.8), the spacetime mass $M$ and the cosmological constant $\Lambda$ can be expressed as:
\begin{eqnarray}
M&=&\frac{3\pi r_{c}^{2}(1+{{x}^{2}})}{16}-\frac{3\pi \Lambda r_{c}^{4}(1+{{x}^{4}})}{96}-\frac{3\pi H(1+x)}{16{{r}_{c}}x}+\frac{{{\pi }^{2}}{{Q}^{2}}(1+{{x}^{2}})}{4r_{c}^{2}{{x}^{2}}}\notag \\
&=&\frac{3\pi r_{c}^{2}{{x}^{2}}}{8(1+{{x}^{2}})}-\frac{3\pi H(1+x+{{x}^{2}}+{{x}^{3}}+{{x}^{4}})}{8{{r}_{c}}x(1+{{x}^{2}})(1+x)}+\frac{{{\pi }^{2}}{{Q}^{2}}(1+{{x}^{2}}+{{x}^{4}})}{2r_{c}^{2}{{x}^{2}}(1+{{x}^{2}})}\label{2.17}\\
\Lambda &=&\frac{6}{r_{c}^{2}(1+{{x}^{2}})}+\frac{6H}{xr_{c}^{5}(1+{{x}^{2}})(1+x)}-\frac{8\pi {{Q}^{2}}}{{{x}^{2}}r_{c}^{6}(1+{{x}^{2}})}\label{2.18}
\end{eqnarray}
here, $x={{r}_{+}}/{{r}_{c}}$.

Substituting Eqs. (2.17) and (2.18) into Eqs. (2.9) and (2.10), we obtain the Hawking radiation temperatures for the black hole horizon (${{T}_{+}}$) and the cosmological horizon (${{T}_{c}}$), respectively, as:
\begin{eqnarray}
&&{{T}_{+}}=\frac{1-{{x}^{2}}}{2\pi {{r}_{c}}x(1+{{x}^{2}})}+\frac{H(1-x)(1+2x+3{{x}^{2}}+4{{x}^{3}})}{4\pi r_{c}^{4}{{x}^{4}}(1+{{x}^{2}})(1+x)}-\frac{2{{Q}^{2}}(1-{{x}^{2}})(1+2{{x}^{2}})}{3{{x}^{5}}r_{c}^{5}(1+{{x}^{2}})}\notag \\
&&{{T}_{c}}=\frac{(1-{{x}^{2}})}{2\pi {{r}_{c}}(1+{{x}^{2}})}+\frac{H(1-x)(4+3x+2{{x}^{2}}+{{x}^{3}})}{4\pi xr_{c}^{4}(1+{{x}^{2}})(1+x)}-\frac{2{{Q}^{2}}(1-{{x}^{2}})(2+{{x}^{2}})}{3{{x}^{2}}r_{c}^{5}(1+{{x}^{2}})}\label{2.19}
\end{eqnarray}

\section{Effective Thermodynamic System} \label{three}

Eqs. (2.11) and (2.14) indicate that the spacetime parameters $M$, $H$, $Q$ and $\Lambda$ correspond to identical thermodynamic state variables at both the black hole and cosmological horizons. However, the thermodynamic quantities associated with each horizon -the entropy ${{S}_{+,c}}$, temperature ${{T}_{+,c}}$, thermodynamic volume ${{V}_{+,c}}$, and electric potential ${{\phi }_{+,c}}$ - generally differ. In particular, the Hawking radiation temperatures ${{T}_{+,c}}$ of the two horizons are typically unequal. Treating the two horizons as independent thermodynamic systems, each with its own temperature, would be an incomplete approach. This is because, in the region where both horizons coexist, the spacetime parameters $M$, $H$, $Q$ and $\Lambda$ are shared. A change in the thermodynamic variables of one horizon inevitably influences those of the other. Therefore, constructing an effective thermodynamic system that captures the global thermodynamic behavior of the coexisting-horizon region is essential for a consistent thermodynamic description.

It is well established that the first law of thermodynamics is a universal relation that any thermodynamic system must satisfy. Consequently, if the parameters $M$, $H$, $Q$ and $\Lambda$ in the spacetime region with coexisting horizons are treated as thermodynamic state variables, the resulting thermodynamic system must obey this fundamental law:
\begin{align}\label{3.1}
dM={{T}_{eff}}dS-{{P}_{eff}}dV+{{\phi }_{eff}}dQ+{{k}_{eff}}dH
\end{align}

The thermodynamic volume of spacetime is \cite{48}:
\begin{align}\label{3.2}
V={{V}_{c}}-{{V}_{+}}=\frac{{{\pi }^{2}}r_{c}^{4}}{2}(1-{{x}^{4}})
\end{align}

Considering dimensional analysis, the total entropy of the system is taken as:
\begin{align}\label{3.3}
S={{S}_{+}}+{{S}_{c}}+\bar{S}=\frac{{{\pi }^{2}}r_{c}^{3}}{2}(1+{{x}^{3}}+f(x))=\frac{{{\pi }^{2}}r_{c}^{3}}{2}F(x)
\end{align}

As illustrated in Figure 2.1, the curve within the coexisting-horizon region consists of segment ABC, representing the black hole horizon, and segment CDE, corresponding to the cosmological horizon. In this region, for fixed values of $H$, $Q$ and $\Lambda$, a single mass value $M$ corresponds to two distinct horizon positions with different Hawking radiation temperatures. The coexistence region terminates at $M={{M}_{\min }}$, where the black hole horizon merges with the inner horizon, resulting in a zero Hawking temperature (${{T}_{+}}=0$). For $M<{{M}_{\min }}$, only the cosmological horizon ${{r}_{c}}$ persists. Consequently, at $M={{M}_{\min }}$, the spacetime exhibits only the cosmological horizon temperature. This endpoint naturally serves as the boundary condition for our effective thermodynamic description of the coexisting region, namely ${{T}_{eff}}({{T}_{+}}=0)={{T}_{c}}({{T}_{+}}=0)$ at $M={{M}_{\min }}$.

From Eq. (3.1) and the boundary conditions, we obtain the effective temperature ${{T}_{eff}}$ and the effective pressure ${{P}_{eff}}$ for the effective system in the coexisting-horizon region as follows:
\begin{eqnarray}
{{T}_{eff}}&=&\frac{(1-{{x}^{2}})}{4\pi {{r}_{c}}{{x}^{6}}}\left[ (1-{{x}^{2}}+{{x}^{4}})+\frac{H[(1+x)(1+{{x}^{2}})(1+{{x}^{5}})-3{{x}^{4}}]}{2r_{c}^{\text{3}}{{x}^{3}}(1+x)}-\frac{4\pi {{Q}^{2}}[(1+{{x}^{2}})(1+{{x}^{6}})-{{x}^{4}}]}{3r_{c}^{\text{4}}{{x}^{4}}} \right]\notag \\
&=&\frac{(1-{{x}^{2}})}{4\pi {{r}_{c}}{{x}^{6}}}\left[ (1-{{x}^{2}}+{{x}^{4}})-\frac{4H{{k}_{eff}}[(1+x)(1+{{x}^{2}})(1+{{x}^{5}})-3{{x}^{4}}](1+{{x}^{2}})}{3\pi r_{c}^{2}{{x}^{2}}(1+x+{{x}^{2}}+{{x}^{3}}+{{x}^{4}})}-\frac{4\pi {{Q}^{2}}[(1+{{x}^{2}})(1+{{x}^{6}})-{{x}^{4}}]}{3r_{c}^{\text{4}}{{x}^{4}}} \right]\notag \\
\label{3.4}
\end{eqnarray}
where ${{k}_{eff}}$ and $H$ represent a pair of conjugate variables \cite{46}.
\begin{eqnarray}
{{P}_{eff}}&=&\frac{3(1-{{x}^{2}})}{8\pi r_{c}^{2}{{x}^{6}}(1+{{x}^{2}})}\left( 1+\frac{H(1+2x+3{{x}^{2}}+4{{x}^{3}})}{2r_{c}^{3}{{x}^{3}}{{(1+x)}^{2}}}-\frac{4\pi {{Q}^{2}}(1+2{{x}^{2}})}{3r_{c}^{4}{{x}^{4}}} \right)F(x)\notag \\
&&-\frac{(1-{{x}^{2}})}{8\pi r_{c}^{2}{{x}^{5}}}\left( 1+\frac{H(1+x+{{x}^{2}}+{{x}^{3}}+{{x}^{4}})}{2r_{c}^{3}{{x}^{3}}(1+x)}-\frac{4\pi {{Q}^{2}}(1+{{x}^{2}}+{{x}^{4}})}{3r_{c}^{4}{{x}^{4}}} \right)F'(x)\label{3.5}
\end{eqnarray}
\begin{align}\label{3.6}
F(x)=\frac{11}{7}{{(1-{{x}^{4}})}^{3/4}}+\frac{3}{7(1-{{x}^{4}})}-1
\end{align}
The effective electrical potential ${{\phi }_{eff}}$ and ${{k}_{eff}}$ for the coexisting-horizon region are:
\begin{align}\label{3.7}
{{\phi }_{eff}}=\frac{{{\pi }^{2}}Q(1+{{x}^{2}}+{{x}^{4}})}{r_{c}^{2}{{x}^{2}}(1+{{x}^{2}})},~~ {{k}_{eff}}=\frac{\partial M}{\partial H}=-\frac{3\pi (1+x+{{x}^{2}}+{{x}^{3}}+{{x}^{4}})}{8{{r}_{c}}x(1+{{x}^{2}})(1+x)}
\end{align}

\section{The heat capacity of effective thermodynamic system} \label{fore}

\subsection*{4.1 In the canonical ensemble}

When the charge $Q$ in spacetime remains constant, the heat capacity of the effective system for the region with coexisting horizons is:
\begin{align}\label{4.1}
{{C}_{Q,\Lambda ,H,{{\kappa }_{eff}}}}={{T}_{eff}}{{\left( \frac{\partial S}{\partial {{T}_{eff}}} \right)}_{Q,\Lambda ,H,{{\kappa }_{eff}}}}
\end{align}

Substituting ${{k}_{eff}}=\frac{\partial M}{\partial H}=-\frac{3\pi (1+x+{{x}^{2}}+{{x}^{3}}+{{x}^{4}})}{8{{r}_{c}}x(1+{{x}^{2}})(1+x)}$ into Eq. (2.18) yields:
\begin{align}\label{4.2}
\Lambda =\frac{6}{r_{c}^{2}(1+{{x}^{2}})}-\frac{16H{{\kappa }_{eff}}}{r_{c}^{4}\pi (1+x+{{x}^{2}}+{{x}^{3}}+{{x}^{4}})}-\frac{8\pi {{Q}^{2}}}{{{x}^{2}}r_{c}^{6}(1+{{x}^{2}})}
\end{align}

From Eq. (4.2), we can solve for:
\begin{align}\label{4.3}
r_{c}^{2}=r_{c}^{2}(\Lambda ,Q,H,{{\kappa }_{eff}},x)
\end{align}

By substituting Eq. (4.3) into (4.1), we plotted the ${{C}_{Q,\Lambda ,H,{{\kappa }_{eff}}}}-{{T}_{eff}}$ and ${{C}_{Q,\Lambda ,H,{{\kappa }_{eff}}}}-x$ curves for different values of the parameters $\Lambda$, $H$, ${{k}_{eff}}$ and $Q$, as shown in Figures 4.1 and 4.2.

\begin{figure}[htbp]
\centering
\subfigure[]{\includegraphics[width=8cm,height=4.5cm]{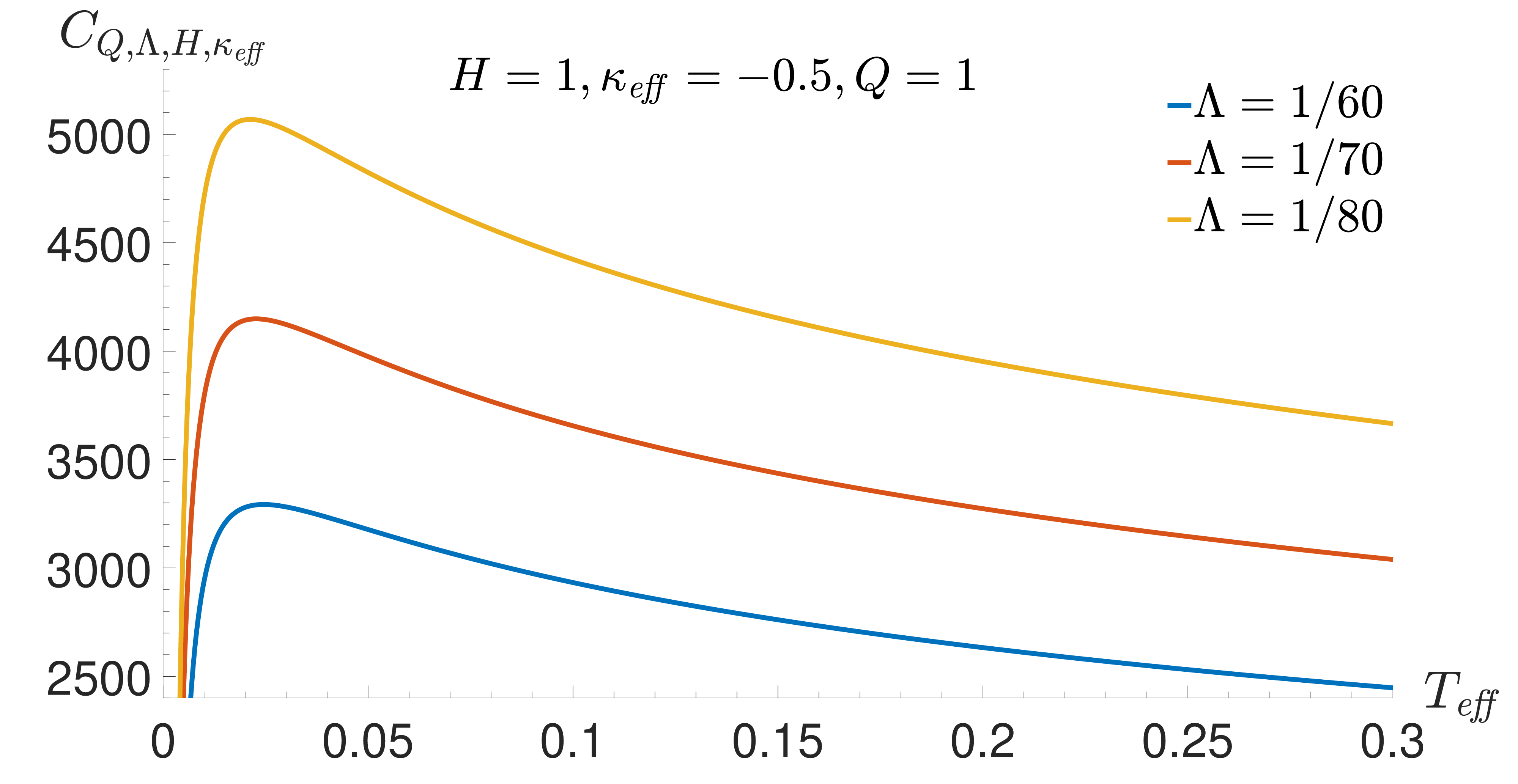}}~
\hspace{0.000001cm}
\subfigure[]{\includegraphics[width=8cm,height=4.5cm]{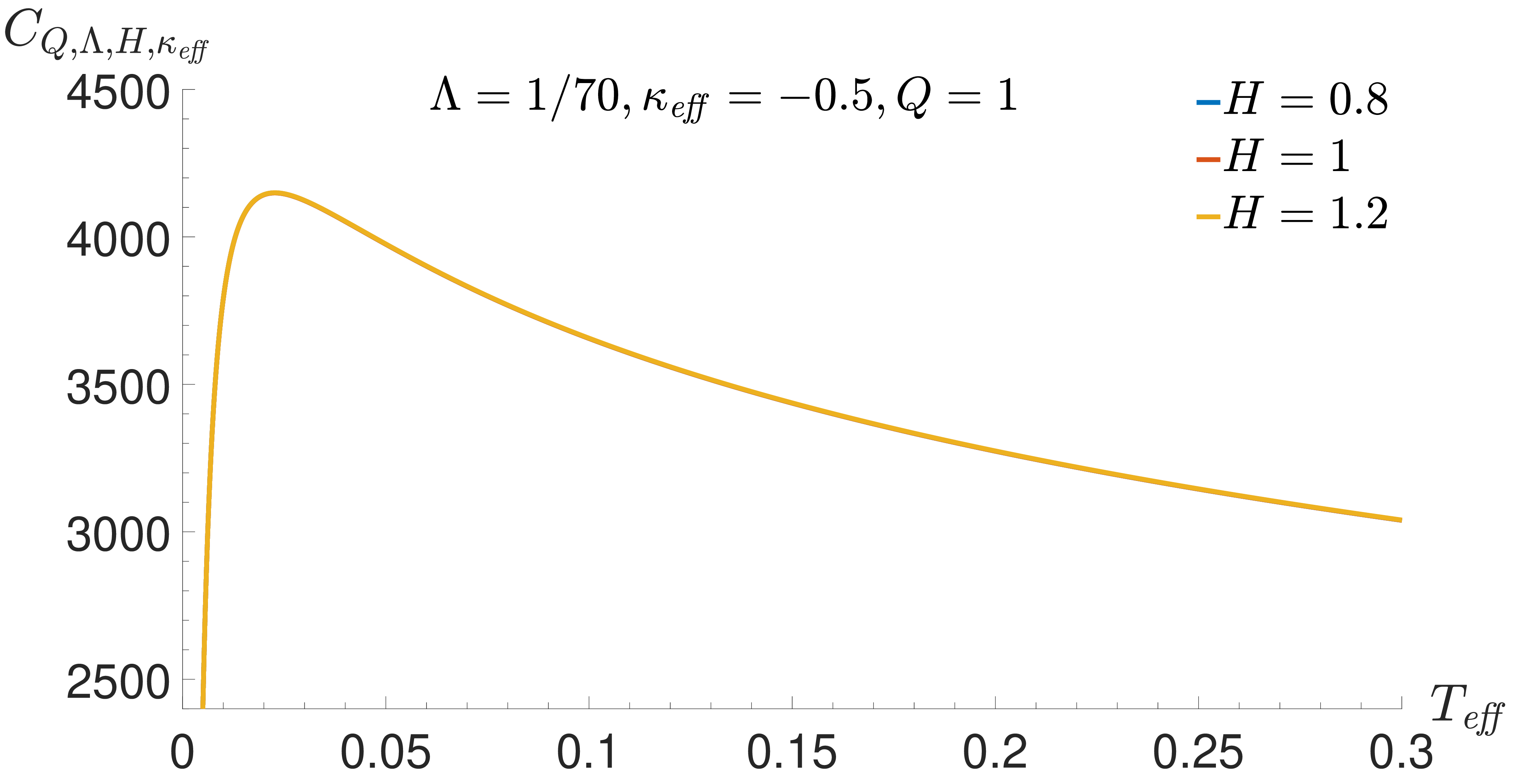}}~
\\
\subfigure[]{\includegraphics[width=8cm,height=4.5cm]{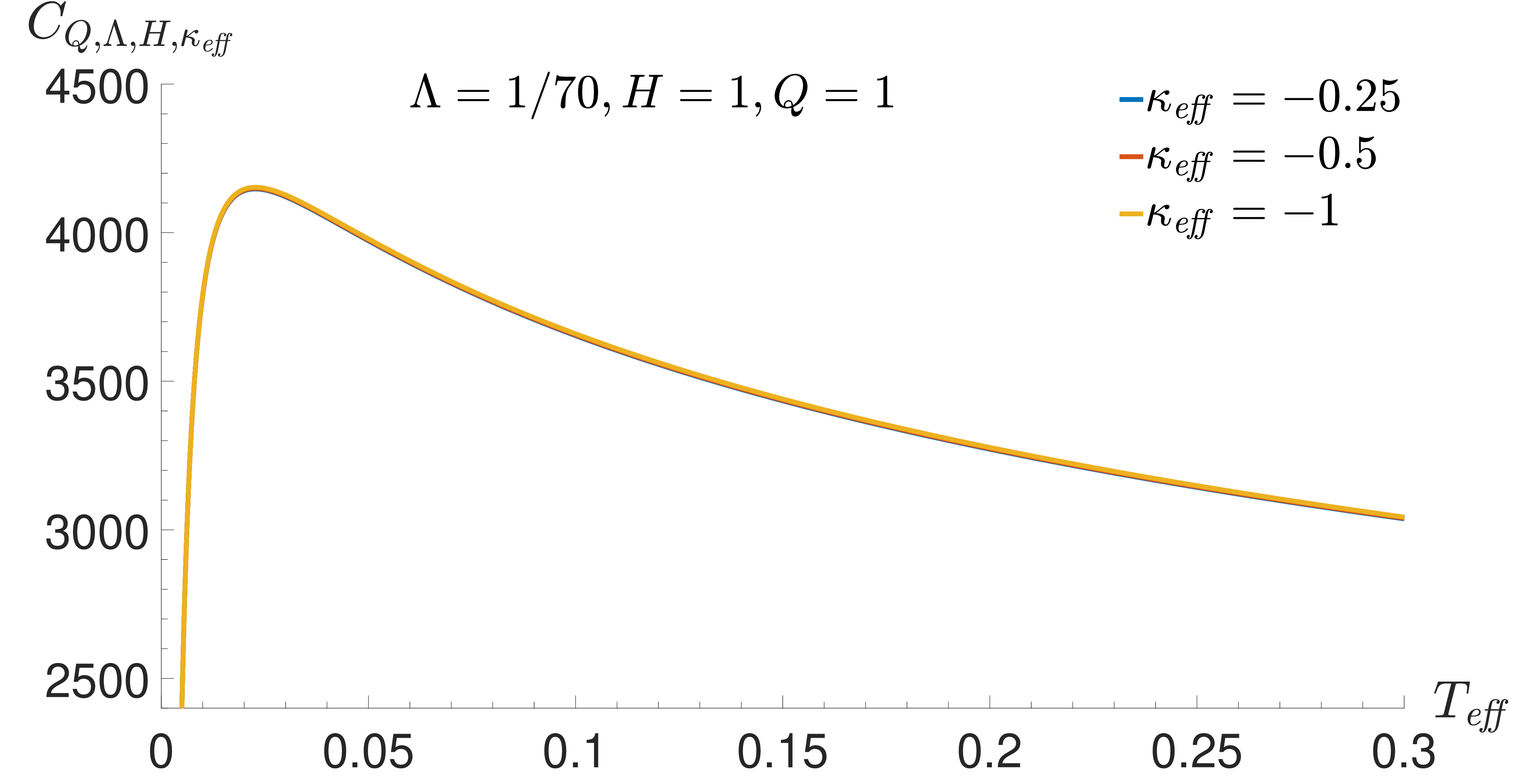}}~
\hspace{0.000001cm}
\subfigure[]{\includegraphics[width=8cm,height=4.5cm]{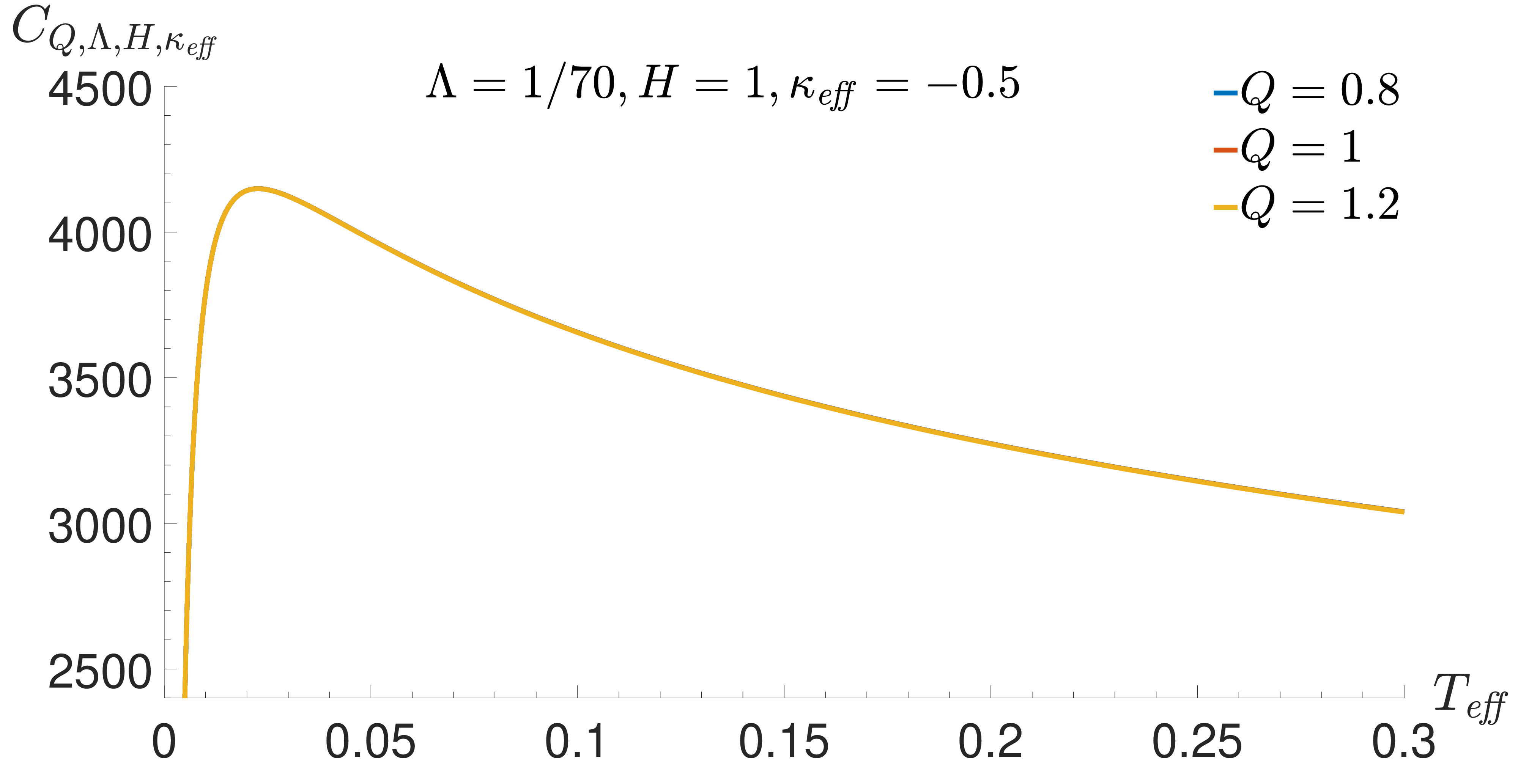}}~
\vskip -1mm \caption{(Color online)The ${{C}_{Q,\Lambda ,H,{{\kappa }_{eff}}}}-{{T}_{eff}}$ curves for different values of the parameters $\Lambda$, $H$, ${{k}_{eff}}$ and $Q$ in the canonical ensemble}
\end{figure}

\begin{figure}[htbp]
\centering
\subfigure[]{\includegraphics[width=8cm,height=4.5cm]{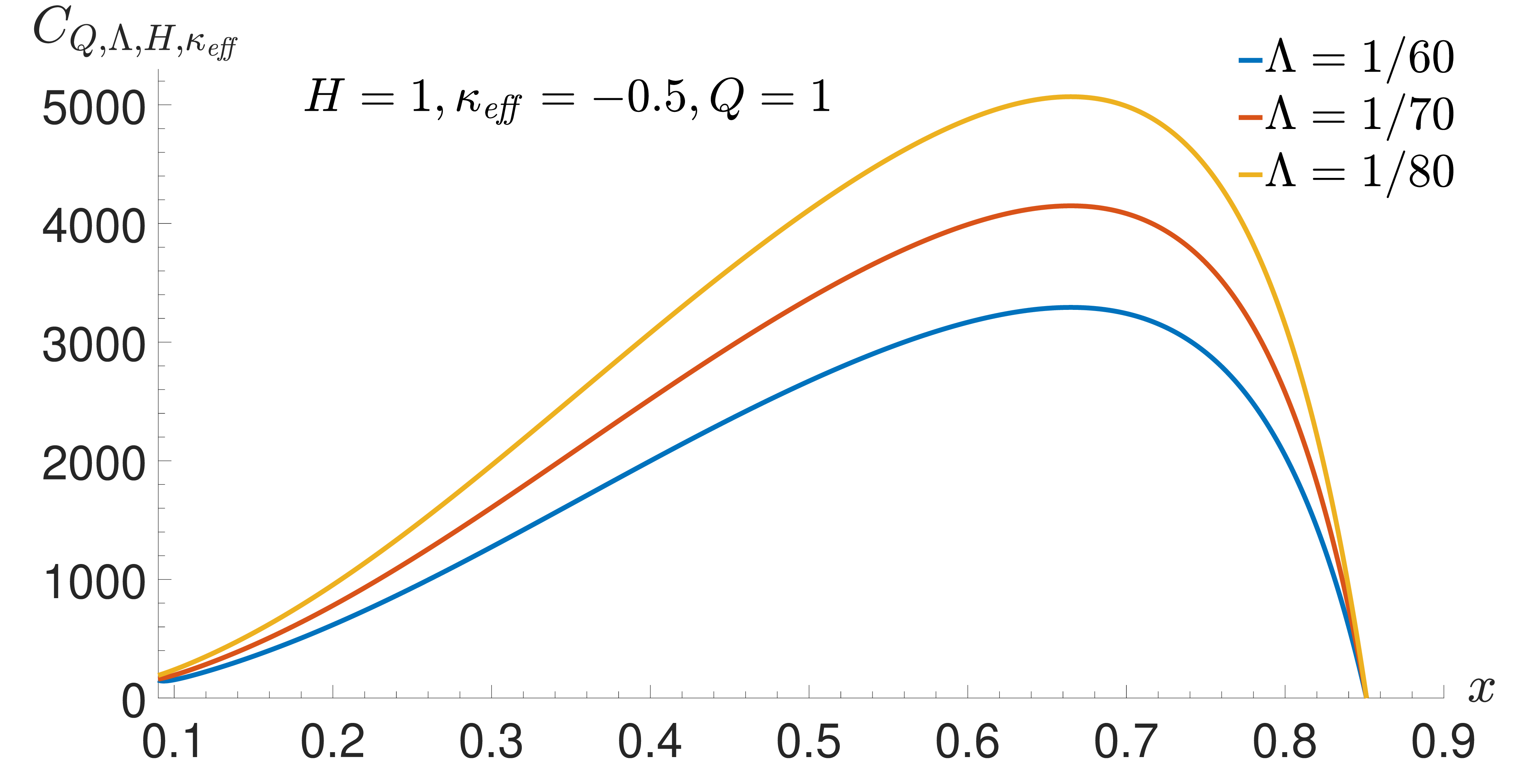}}~
\hspace{0.000001cm}
\subfigure[]{\includegraphics[width=8cm,height=4.5cm]{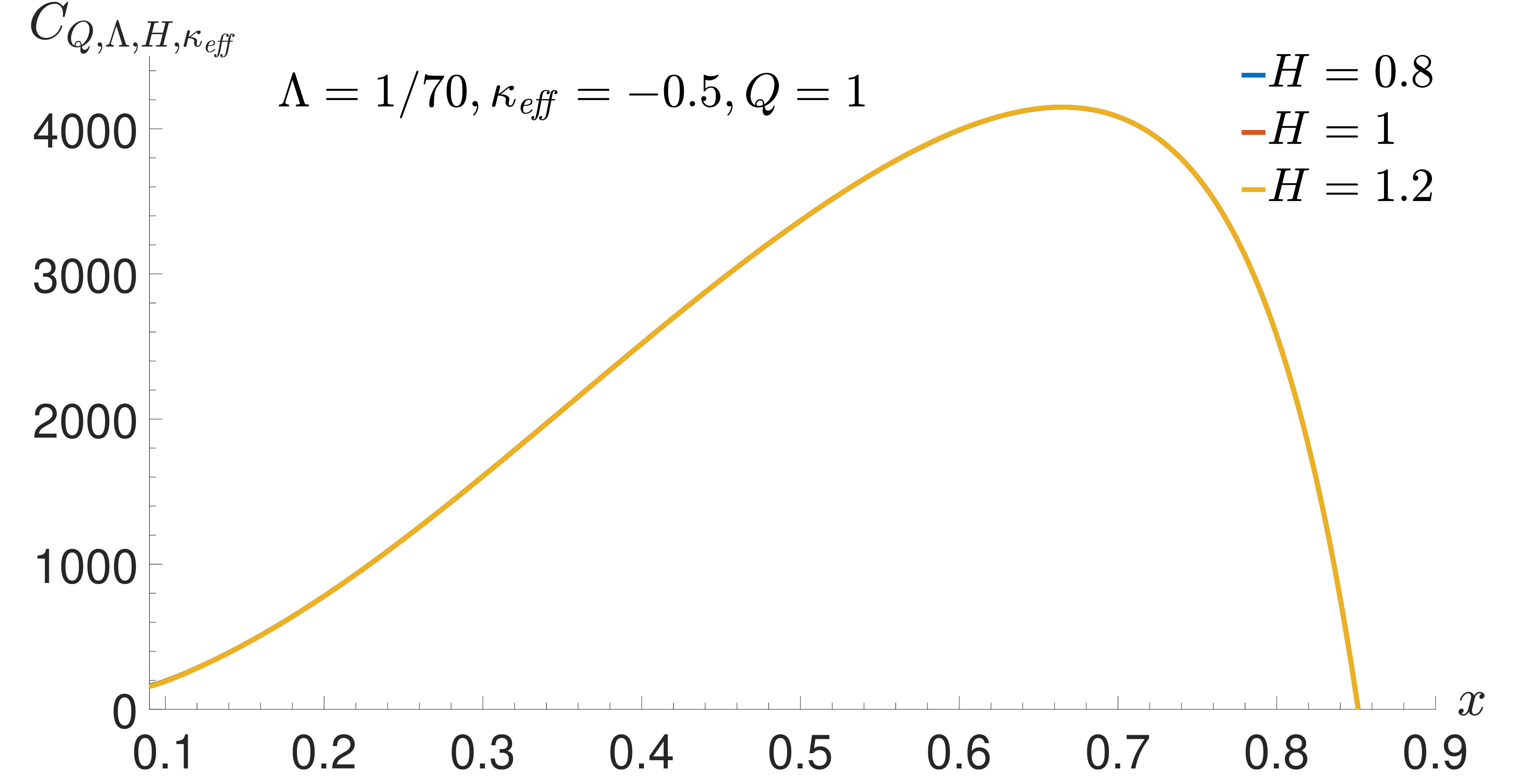}}~
\\
\subfigure[]{\includegraphics[width=8cm,height=4.5cm]{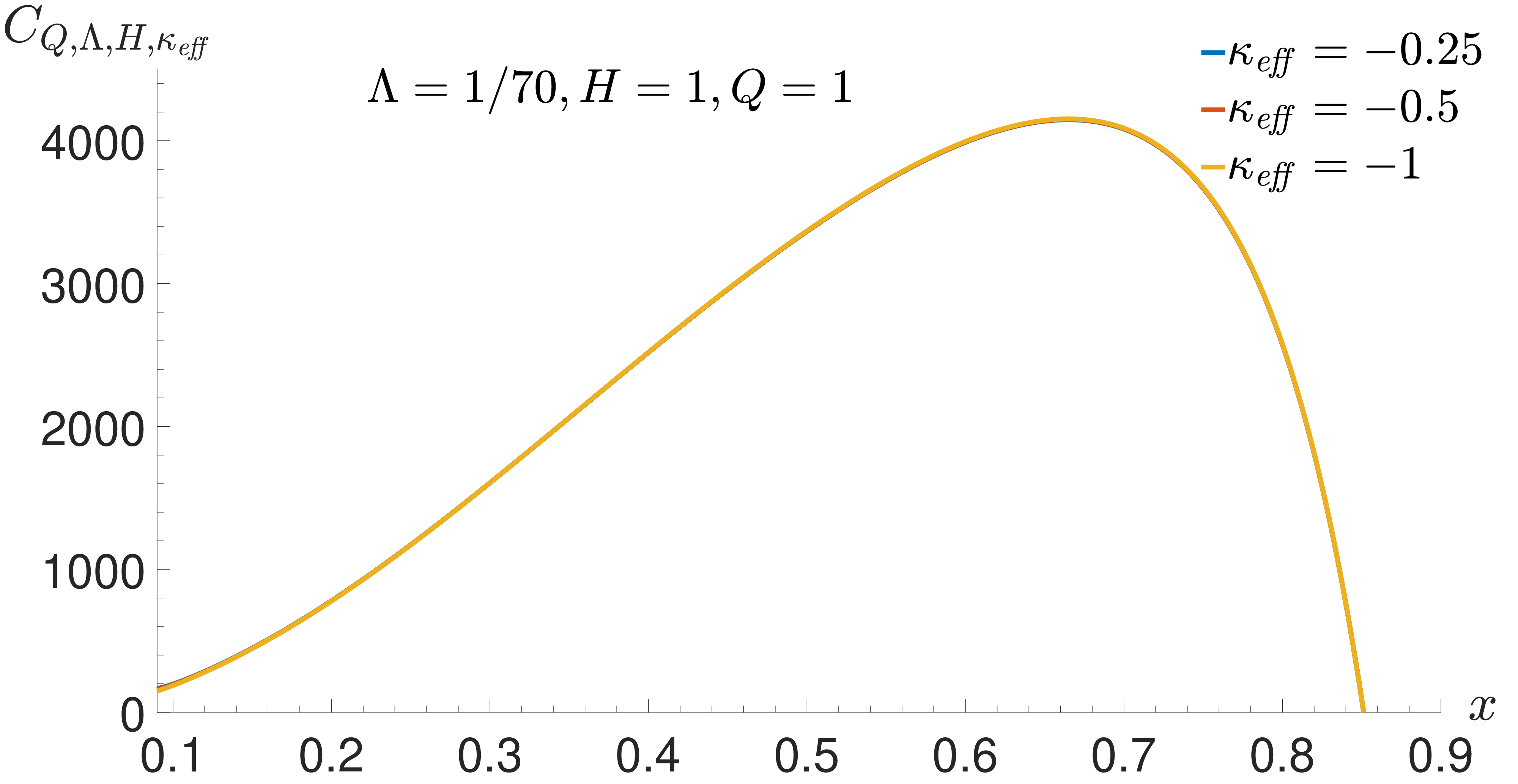}}~
\hspace{0.000001cm}
\subfigure[]{\includegraphics[width=8cm,height=4.5cm]{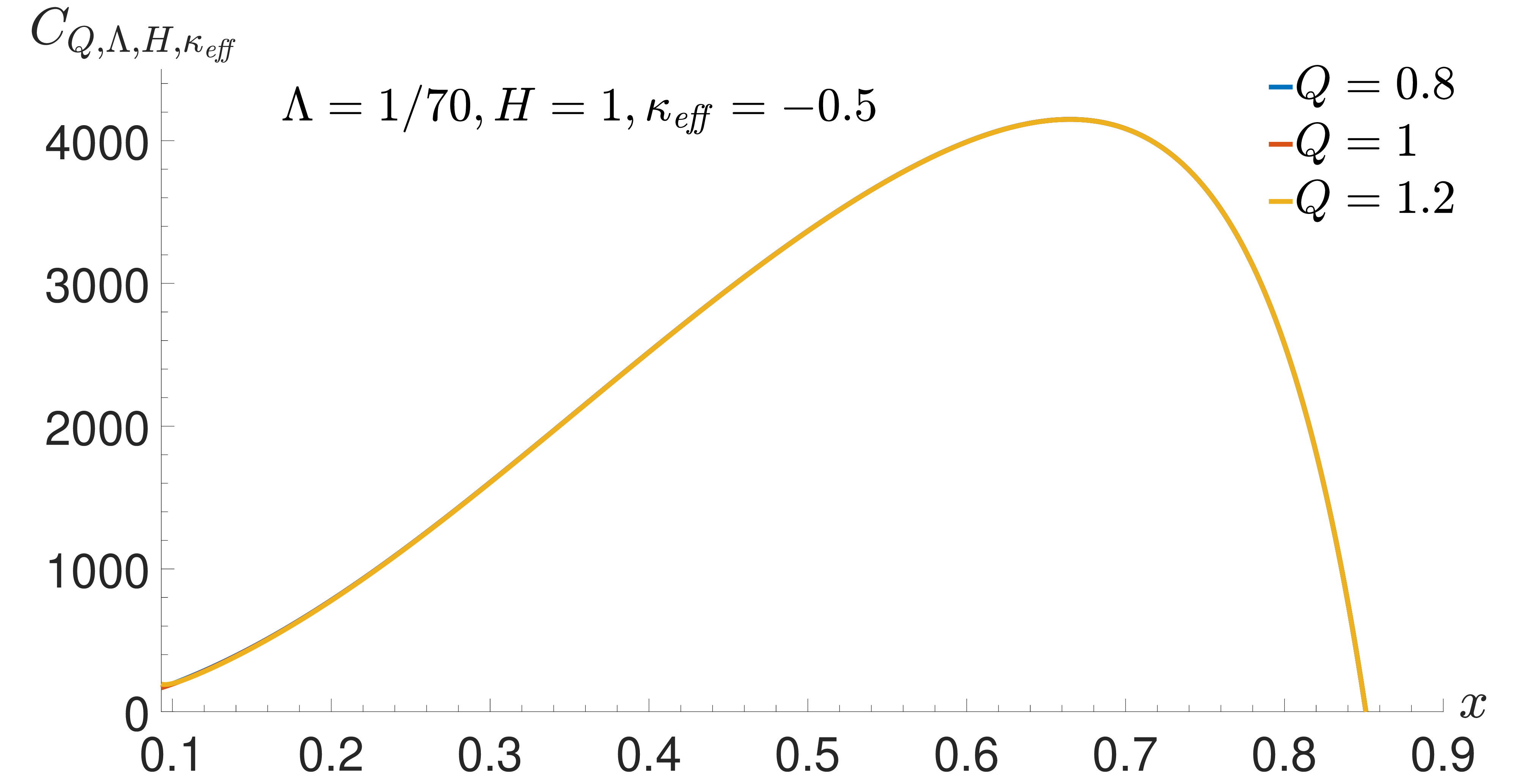}}~
\vskip -1mm \caption{(Color online)The ${{C}_{Q,\Lambda ,H,{{\kappa }_{eff}}}}-x$ curves for different values of the parameters $\Lambda$, $H$, ${{k}_{eff}}$ and $Q$ in the canonical ensemble}
\end{figure}

As shown in Figures 4.1 and 4.2, the ${{C}_{Q,\Lambda ,H,{{\kappa }_{eff}}}}-{{T}_{eff}}$ and ${{C}_{Q,\Lambda ,H,{{\kappa }_{eff}}}}-x$ curves of the canonical ensemble exhibit variations under different constraints. Nevertheless, all curves share two prominent universal features: 1. The heat capacity ${{C}_{Q,\Lambda ,H,{{\kappa }_{eff}}}}$ exhibits a distinct maximum as a function of either temperature ${{T}_{eff}}$ or the horizon position ratio $x$ ; 2. Under all constraint conditions, the heat capacity ${{C}_{Q,\Lambda ,H,{{\kappa }_{eff}}}}\to 0$ as ${{T}_{eff}}\to 0$ (or as $x\to 1$).This indicates that the heat capacity ${{C}_{Q,\Lambda ,H,{{\kappa }_{eff}}}}$ in the two-horizon regime of the 5-dimensional de Sitter hairy spacetime, whether as a function of temperature ${{T}_{eff}}$ or the ratio $x$, exhibits characteristics analogous to the temperature-dependent heat capacity of a two-level system \cite{49,50,51}.

Furthermore, Figures 4.1 and 4.2 reveal that the values of parameters $H$, ${{k}_{eff}}$ and $Q$ have negligible impact on the curves. In contrast, the value of the cosmological constant $\Lambda$ significantly influences the magnitude of the maximum in the heat capacity curves.

\subsection*{4.2 In the grand canonical ensemble}

When the effective electric potential ${{\phi }_{eff}}$ in spacetime remains constant, substituting $\frac{Q}{r_{c}^{2}}=\frac{{{\phi }_{eff}}{{x}^{2}}(1+{{x}^{2}})}{{{\pi }^{2}}(1+{{x}^{2}}+{{x}^{4}})}$ into Eq. (4.2) yields:
\begin{align}\label{4.4}
r_{c}^{4}-\frac{2}{(1+{{x}^{2}})\Lambda }\left( 3-\frac{4\pi }{{{x}^{2}}}{{\left( \frac{{{\phi }_{eff}}{{x}^{2}}(1+{{x}^{2}})}{{{\pi }^{2}}(1+{{x}^{2}}+{{x}^{4}})} \right)}^{2}} \right)r_{c}^{2}+\frac{16H{{\kappa }_{eff}}}{\pi \Lambda (1+x+{{x}^{2}}+{{x}^{3}}+{{x}^{4}})}=0
\end{align}

Solving Eq. (4.4) gives:
\begin{align}\label{4.5}
r_{c}^{2}=\frac{-\bar{b}\pm \sqrt{{{{\bar{b}}}^{2}}-4\bar{c}}}{2}
\end{align}
where
\begin{align}\label{4.6}
\bar{b}=-\frac{2}{(1+{{x}^{2}})\Lambda }\left( 3-\frac{4\pi }{{{x}^{2}}}{{\left( \frac{{{\phi }_{eff}}{{x}^{2}}(1+{{x}^{2}})}{{{\pi }^{2}}(1+{{x}^{2}}+{{x}^{4}})} \right)}^{2}} \right), \bar{c}=\frac{16H{{\kappa }_{eff}}}{\pi \Lambda (1+x+{{x}^{2}}+{{x}^{3}}+{{x}^{4}})}
\end{align}

By substituting Eq. (4.5) into (4.1), we plotted the ${{C}_{{{\phi }_{eff}},\Lambda ,H,{{\kappa }_{eff}}}}-{{T}_{eff}}$ and ${{C}_{{{\phi }_{eff}},\Lambda ,H,{{\kappa }_{eff}}}}-x$ curves for different values of the parameters $\Lambda$, $H$, ${{k}_{eff}}$ and ${{\phi }_{eff}}$, as shown in Figures 4.3 and 4.4.

\begin{figure}[htbp]
\centering
\subfigure[]{\includegraphics[width=8cm,height=4.5cm]{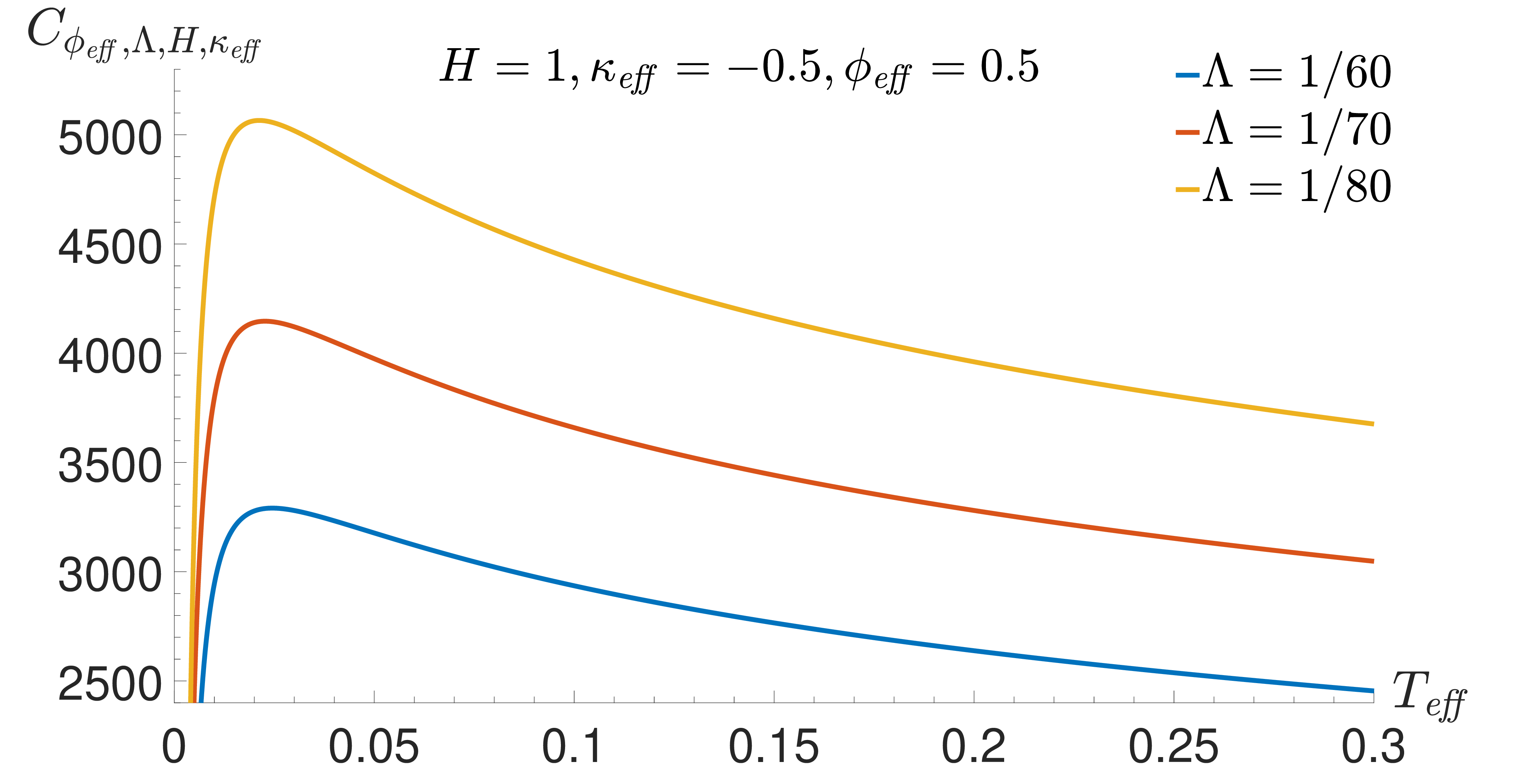}}~
\hspace{0.000001cm}
\subfigure[]{\includegraphics[width=8cm,height=4.5cm]{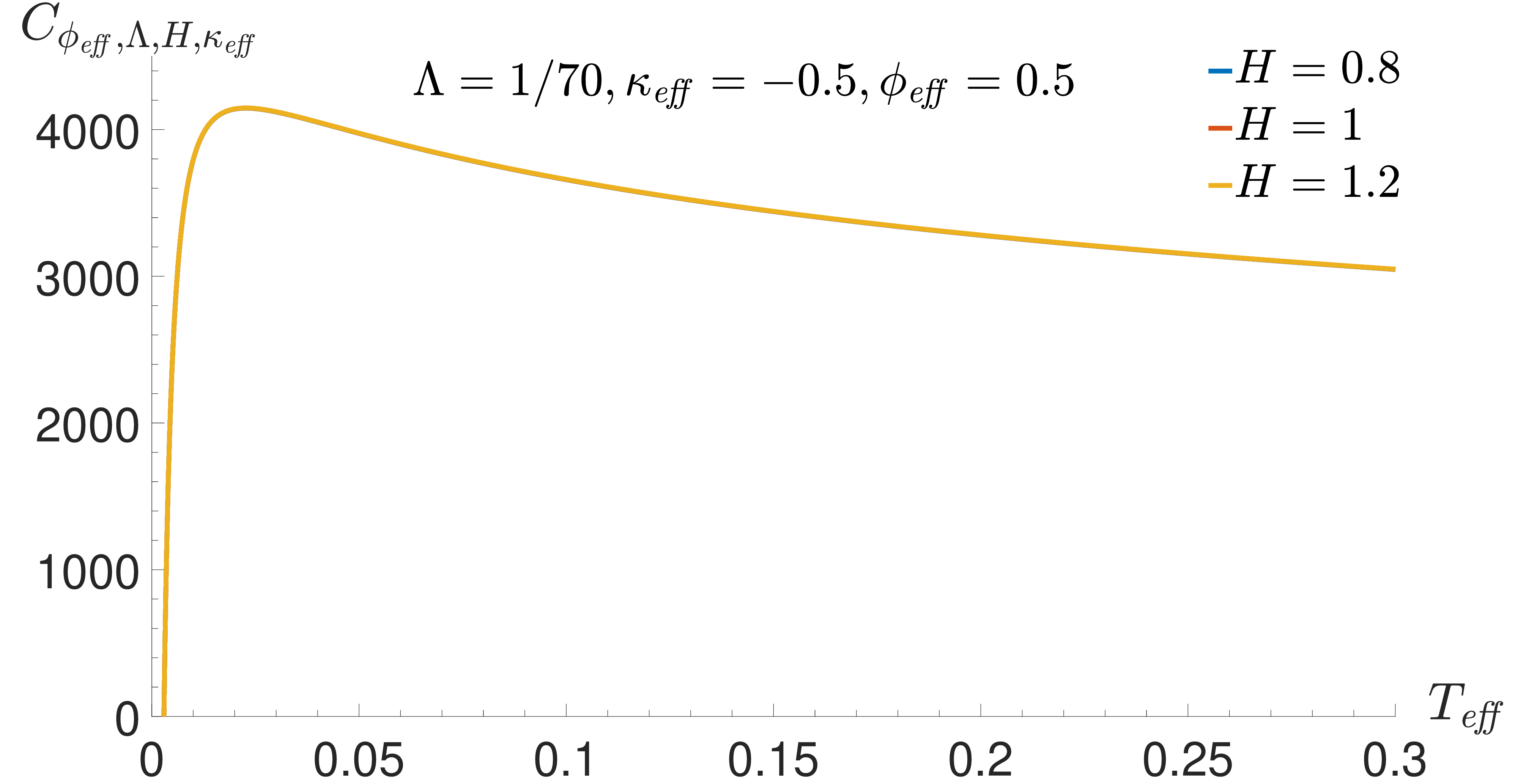}}~
\\
\subfigure[]{\includegraphics[width=8cm,height=4.5cm]{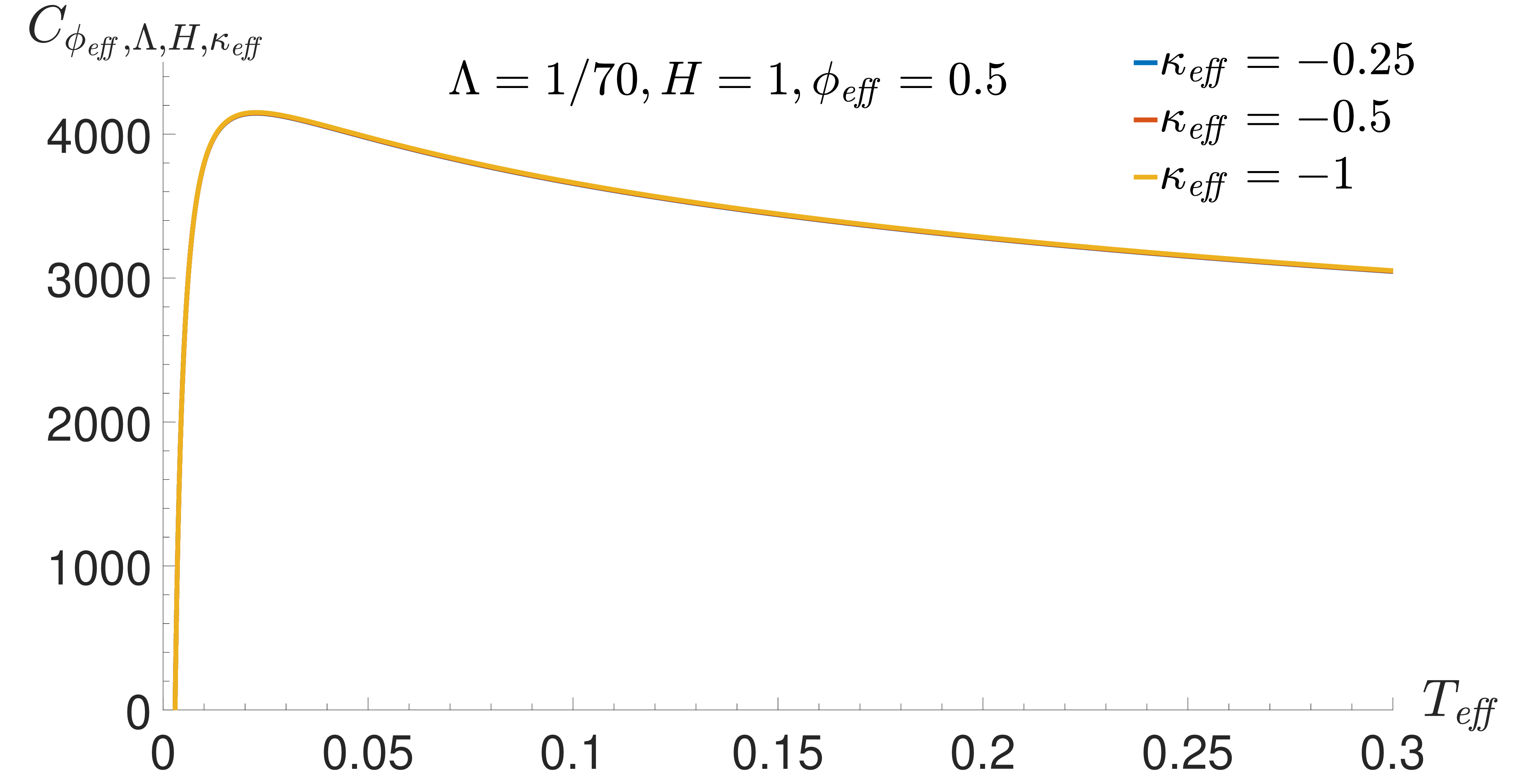}}~
\hspace{0.000001cm}
\subfigure[]{\includegraphics[width=8cm,height=4.5cm]{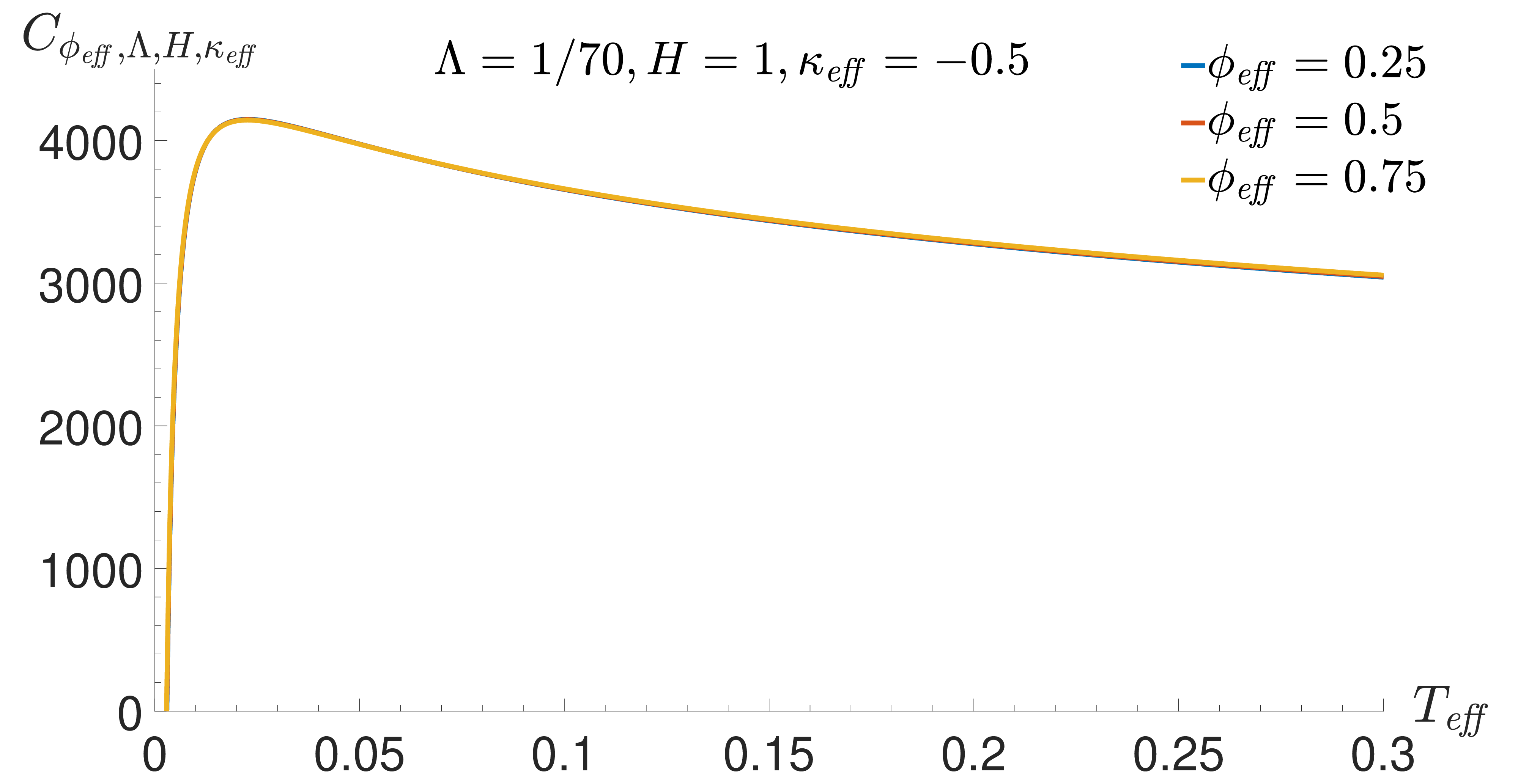}}~
\vskip -1mm \caption{(Color online)The ${{C}_{{{\phi }_{eff}},\Lambda ,H,{{\kappa }_{eff}}}}-{{T}_{eff}}$ curves for different values of the parameters $\Lambda$, $H$, ${{k}_{eff}}$ and ${{\phi }_{eff}}$ in the grand canonical ensemble}
\end{figure}

\begin{figure}[htbp]
\centering
\subfigure[]{\includegraphics[width=8cm,height=4.5cm]{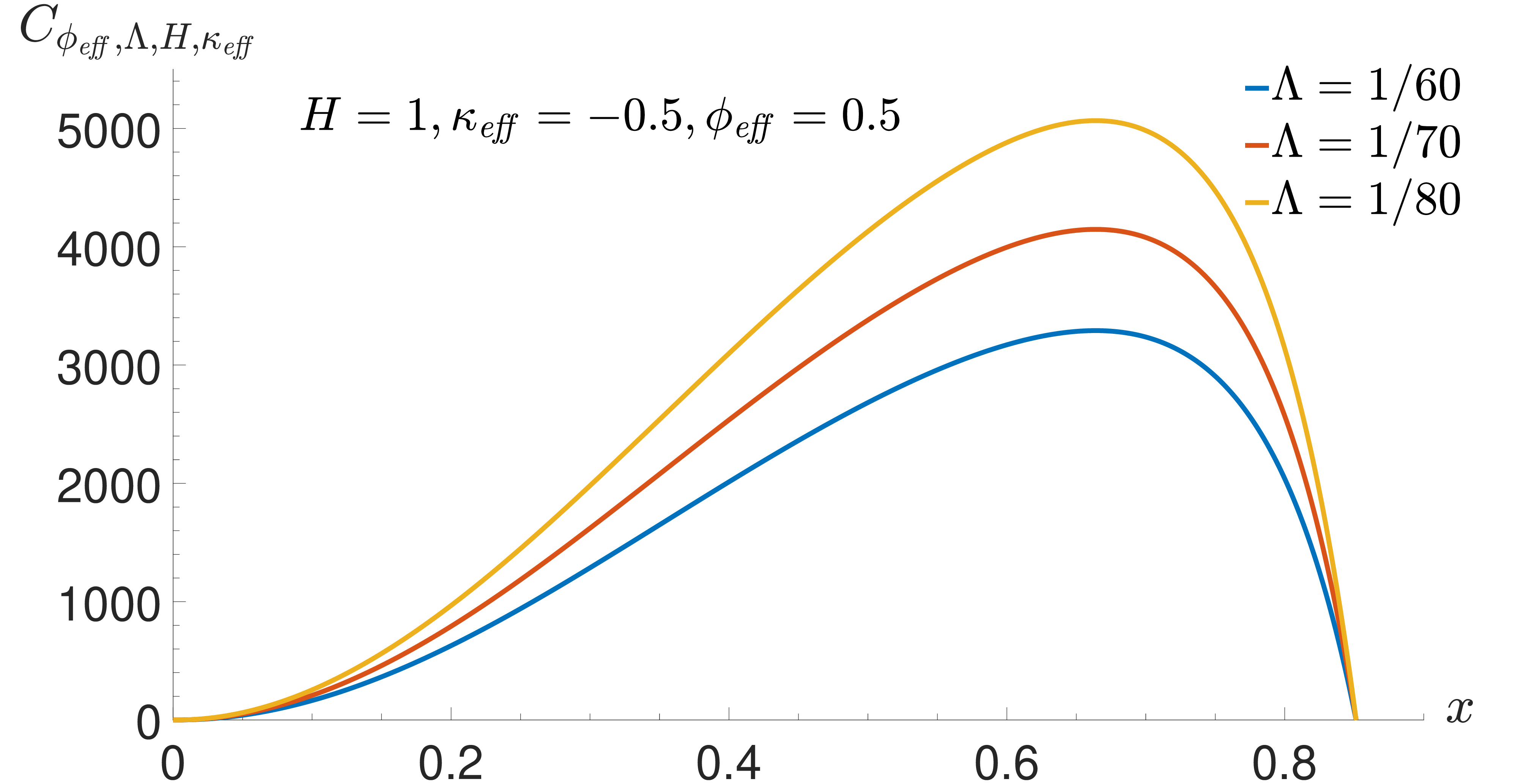}}~
\hspace{0.000001cm}
\subfigure[]{\includegraphics[width=8cm,height=4.5cm]{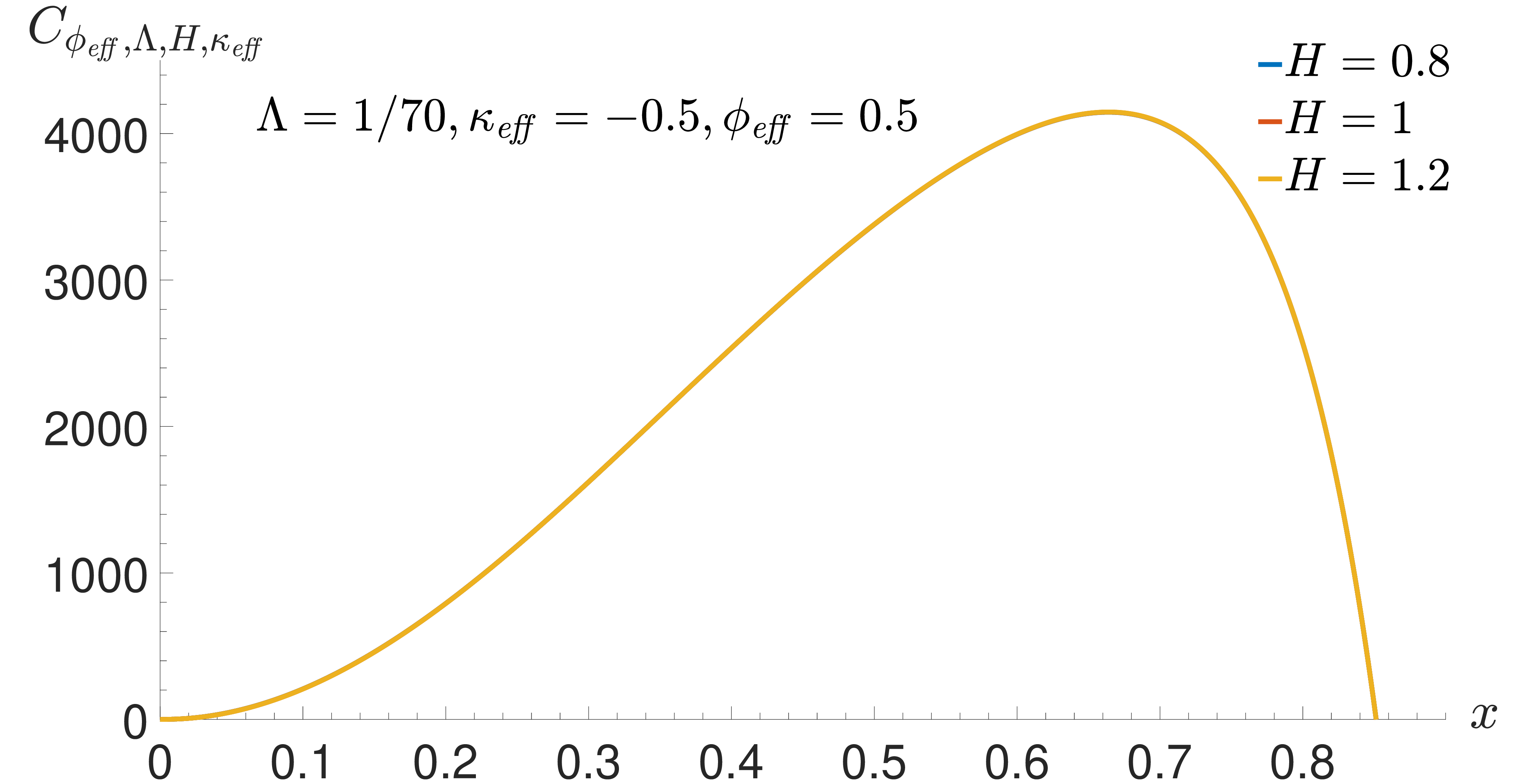}}~
\\
\subfigure[]{\includegraphics[width=8cm,height=4.5cm]{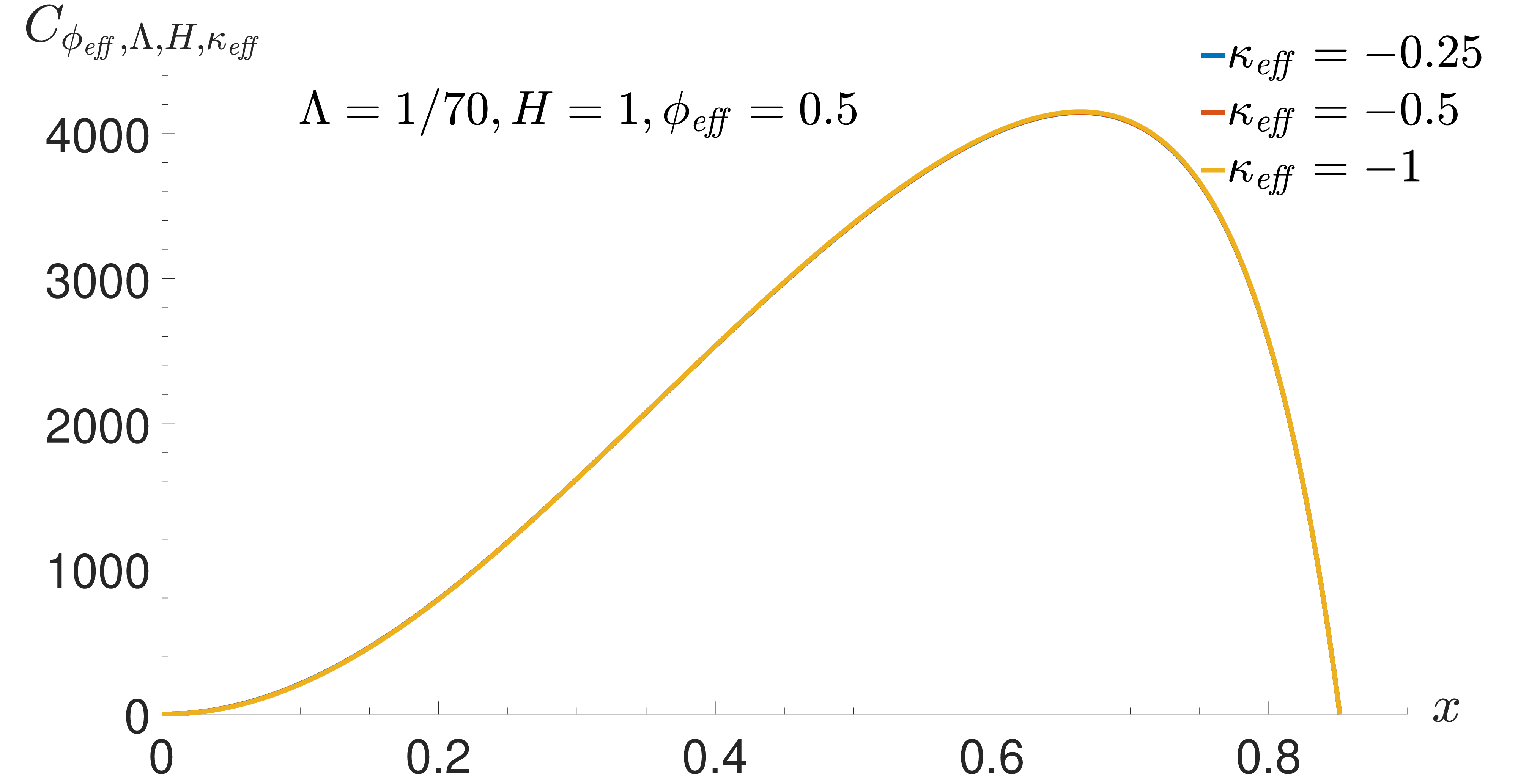}}~
\hspace{0.000001cm}
\subfigure[]{\includegraphics[width=8cm,height=4.5cm]{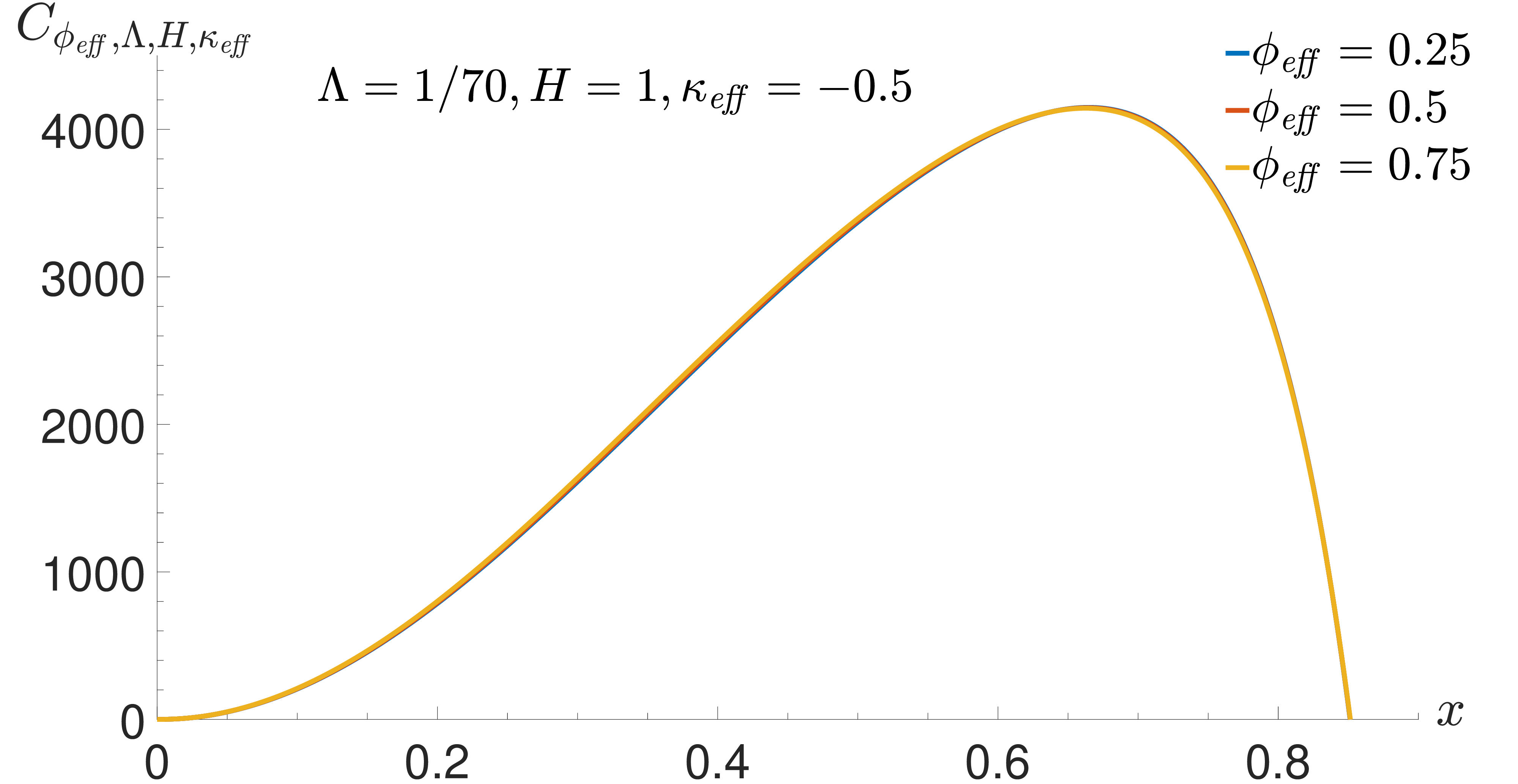}}~
\vskip -1mm \caption{(Color online)The ${{C}_{{{\phi }_{eff}},\Lambda ,H,{{\kappa }_{eff}}}}-x$ curves for different values of the parameters $\Lambda$, $H$, ${{k}_{eff}}$ and ${{\phi }_{eff}}$ in the grand canonical ensemble}
\end{figure}

As illustrated in Figures 4.3 and 4.4, the ${{C}_{{{\phi }_{eff}},\Lambda ,H,{{\kappa }_{eff}}}}-{{T}_{eff}}$ and ${{C}_{{{\phi }_{eff}},\Lambda ,H,{{\kappa }_{eff}}}}-x$ curves of the grand canonical ensemble, under various constraints, exhibit characteristics analogous to those of the canonical ensemble. The parameters ${{k}_{eff}}$, $H$ and ${{\phi }_{eff}}$ are found to have a negligible influence on the curves, with only the cosmological constant $\Lambda$ significantly affecting the maximum value. This comparative result demonstrates that the distinctive variation in heat capacity within the coexisting-horizon region of the 5-dimensional de Sitter hairy spacetime is intrinsic to the system itself, rather than being dictated by the specific constraint conditions applied.

This finding raises a profound question: are the thermodynamic properties of the two-horizon coexistent regime in this spacetime fundamentally linked to those of a two-level thermodynamic system? In the following section, we will delve into this potential connection.

\section{Schottky anomaly of the 5-dimensional de Sitter hairy spacetime}\label{five}

The heat capacity of a thermal system is typically an increasing function of temperature. However, a peak in the heat capacity, known as a Shottky anomaly can occur in a system that has a maximum energy. The partition function for a two-level paramagnetic system is:
\begin{align}\label{5.1}
z={{e}^{\beta \varepsilon }}+{{e}^{-\beta \varepsilon }}
\end{align}

The thermodynamic function is obtained from Eq. (5.1):
\begin{eqnarray}
S(T,H,N)&=&Nk[\ln 2\cosh (\beta \varepsilon )-\beta \varepsilon \tanh (\beta \varepsilon )]\notag \\
U&=&F+TS=-Nk\tanh (\beta \varepsilon )\notag \\
{{C}_{H}}&=&{{\left( \frac{\partial U}{\partial T} \right)}_{H}}=Nk{{(\beta \varepsilon )}^{2}}{{\cosh }^{-2}}(\beta \varepsilon )\label{5.2}
\end{eqnarray}
where $\frac{1}{\beta \varepsilon }=\frac{kT}{\varepsilon }$, $k$ is the Boltzmann constant($k=1.381\times {{10}^{-23}}J\cdot {{K}^{-1}}$), $N$ is the number of microscopic particles in the two-level system. When the interactions between particles are neglected, the heat capacity assumes the following universal form:
\begin{align}\label{5.3}
C=Nk{{\left( \frac{\Delta }{T} \right)}^{2}}\frac{{{e}^{\tfrac{\Delta }{T}}}}{{{(1+{{e}^{\tfrac{\Delta }{T}}})}^{2}}}
\end{align}
where the energy gap $\Delta =2\varepsilon $. This behavior is known as the Schottky specific heat. As such, it is of great importance because it provides a universal model for a wide variety of systems that can be approximated as two-level systems \cite{51}. The corresponding curve of energy and specific heat as functions of temperature are depicted in Figure 5.1.
\begin{figure}[htb]
\centering
\includegraphics[width=8cm,height=4.5cm]{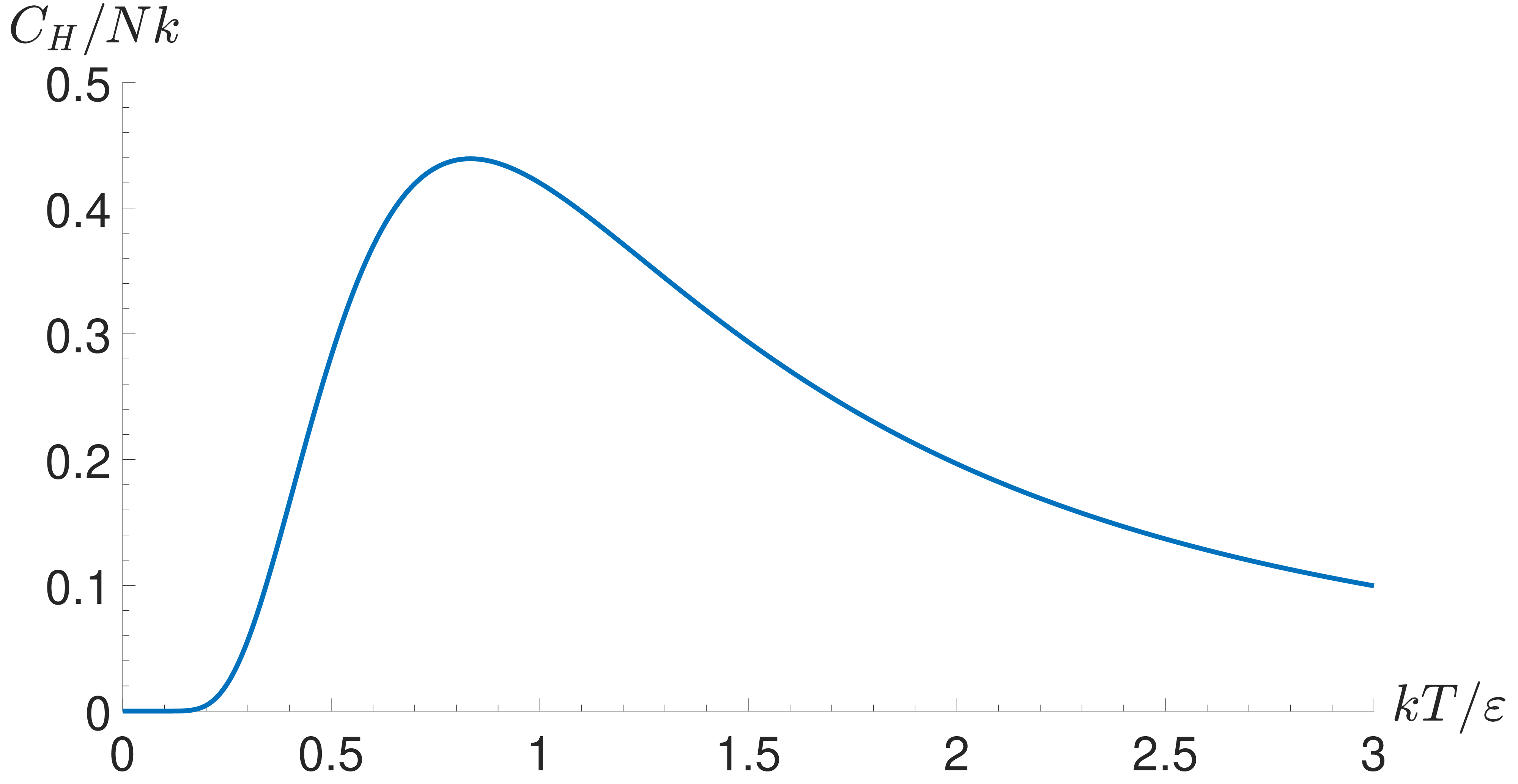}
\caption{(Color online) The $\frac{kT}{\varepsilon }-\frac{{{C}_{H}}}{Nk}$ curve}
\end{figure}

As shown in Figure 5.1, a quantum system with a split spectrum exhibits the characteristic that $C\to 0$ as $T\to 0$. Consequently, the specific heat displays a maximum at approximately $\varepsilon \approx kT$, which is a hallmark feature of a two-level system with an energy gap of $\Delta =2\varepsilon $.

As derived from the analysis in Section 3, the effective thermodynamic system consists of two horizons with distinct radiation temperatures. A key issue in studying the thermodynamic properties of dS spacetime is to understand the thermal characteristics between these two horizons. Recent studies have made some progress in investigating the Schottky-specific heat in dS spacetime \cite{49,50,51}. However, these discussions did not account for the correlation between the black hole horizon and the cosmological horizon, leading to incomplete conclusions. In this section, we analyze the Schottky-specific heat in the coexisting region of the two horizons within a 5-dimensional de Sitter hairy spacetime, taking into consideration their mutual interaction. Since the radiation temperatures on the black hole horizon and the cosmological horizon in this 5-dimensional de Sitter hairy spacetime differ, we treat the two horizons as distinct energy levels. When interparticle interactions are neglected, particles residing on different horizons occupy different energy levels, and the heat capacity of particles transitioning between these two energy levels is given by:

\begin{align}\label{5.4}
{{\hat{C}}_{Q,\Lambda ,H,{{\kappa }_{eff}}}}=Nk{{\left( \frac{\varepsilon \Delta }{{{T}_{eff}}} \right)}^{2}}\frac{{{e}^{\tfrac{\varepsilon \Delta }{{{T}_{eff}}}}}}{{{(1+{{e}^{\tfrac{\varepsilon \Delta }{{{T}_{eff}}}}})}^{2}}}
\end{align}
here
\begin{align}\label{5.5}
\frac{\varepsilon \Delta }{{{T}_{eff}}}=\frac{{{T}_{+}}-{{T}_{c}}}{{{T}_{eff}}}
\end{align}

For canonical ensemble, By substituting Eqs. (2.19) and (3.4) into (5.4), we plotted the ${{\hat{C}}_{Q,\Lambda ,H,{{\kappa }_{eff}}}}/Nk-{{T}_{eff}}$ and ${{\hat{C}}_{Q,\Lambda ,H,{{\kappa }_{eff}}}}/Nk-x$ curves for different values of the parameters $\Lambda$, $H$, ${{k}_{eff}}$ and $Q$, as shown in Figures 5.2 and 5.3.

\begin{figure}[htbp]
\centering
\subfigure[]{\includegraphics[width=8cm,height=4.5cm]{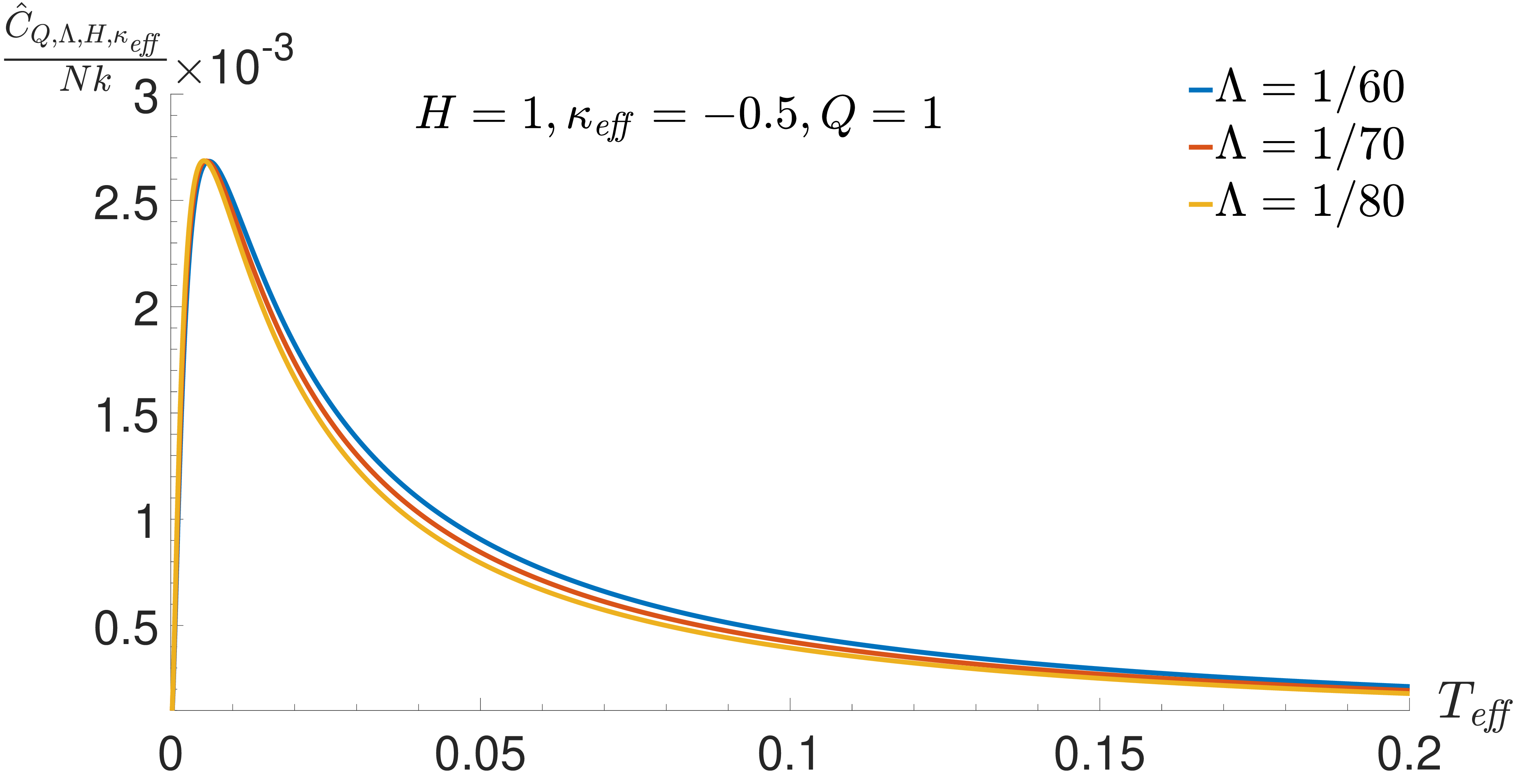}}~
\hspace{0.000001cm}
\subfigure[]{\includegraphics[width=8cm,height=4.5cm]{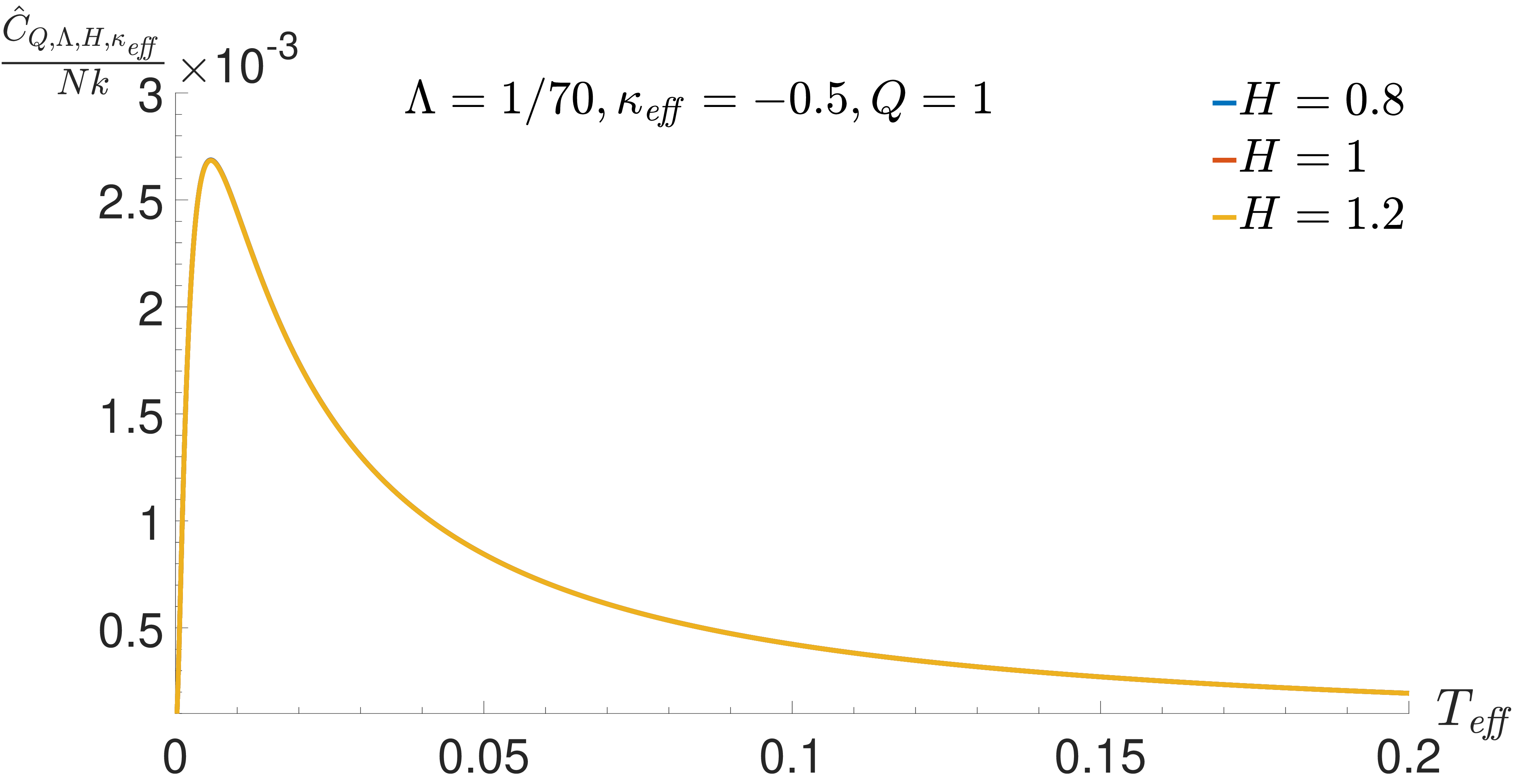}}~
\\
\subfigure[]{\includegraphics[width=8cm,height=4.5cm]{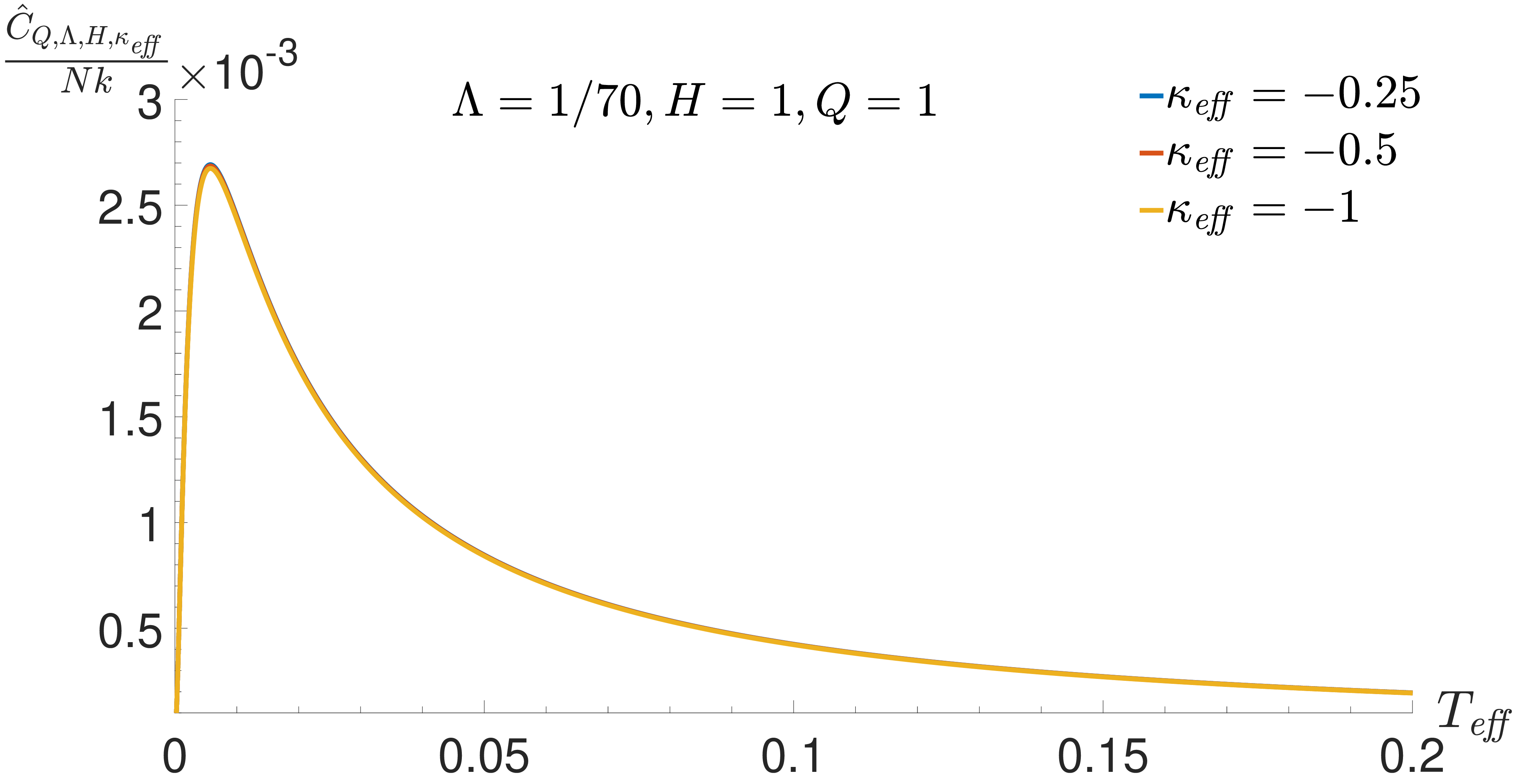}}~
\hspace{0.000001cm}
\subfigure[]{\includegraphics[width=8cm,height=4.5cm]{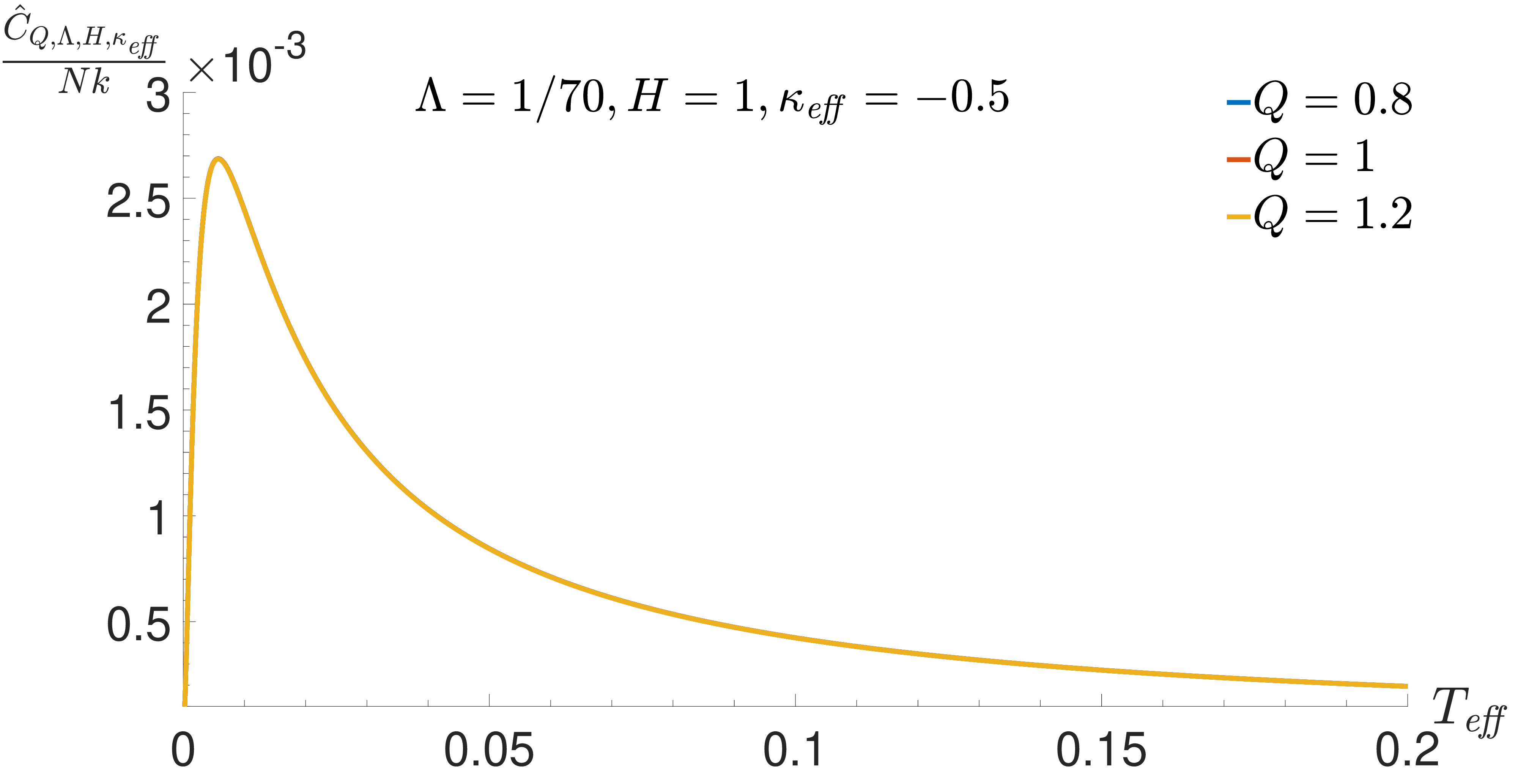}}~
\vskip -1mm \caption{(Color online)The ${{\hat{C}}_{Q,\Lambda ,H,{{\kappa }_{eff}}}}/Nk-{{T}_{eff}}$ curves for different values of the parameters $\Lambda$, $H$, ${{k}_{eff}}$ and $Q$ in the canonical ensemble}
\end{figure}

\begin{figure}[htbp]
\centering
\subfigure[]{\includegraphics[width=8cm,height=4.5cm]{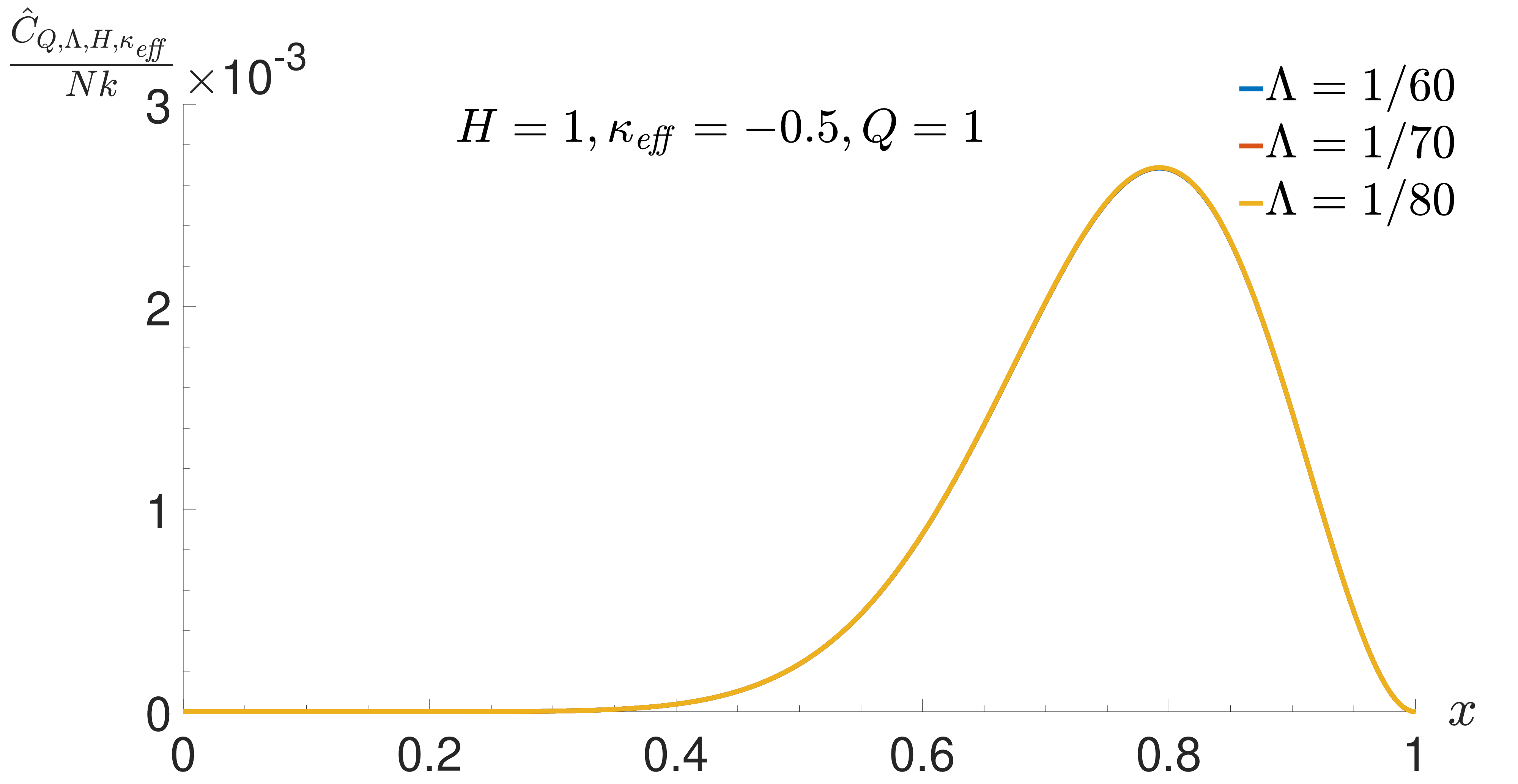}}~
\hspace{0.000001cm}
\subfigure[]{\includegraphics[width=8cm,height=4.5cm]{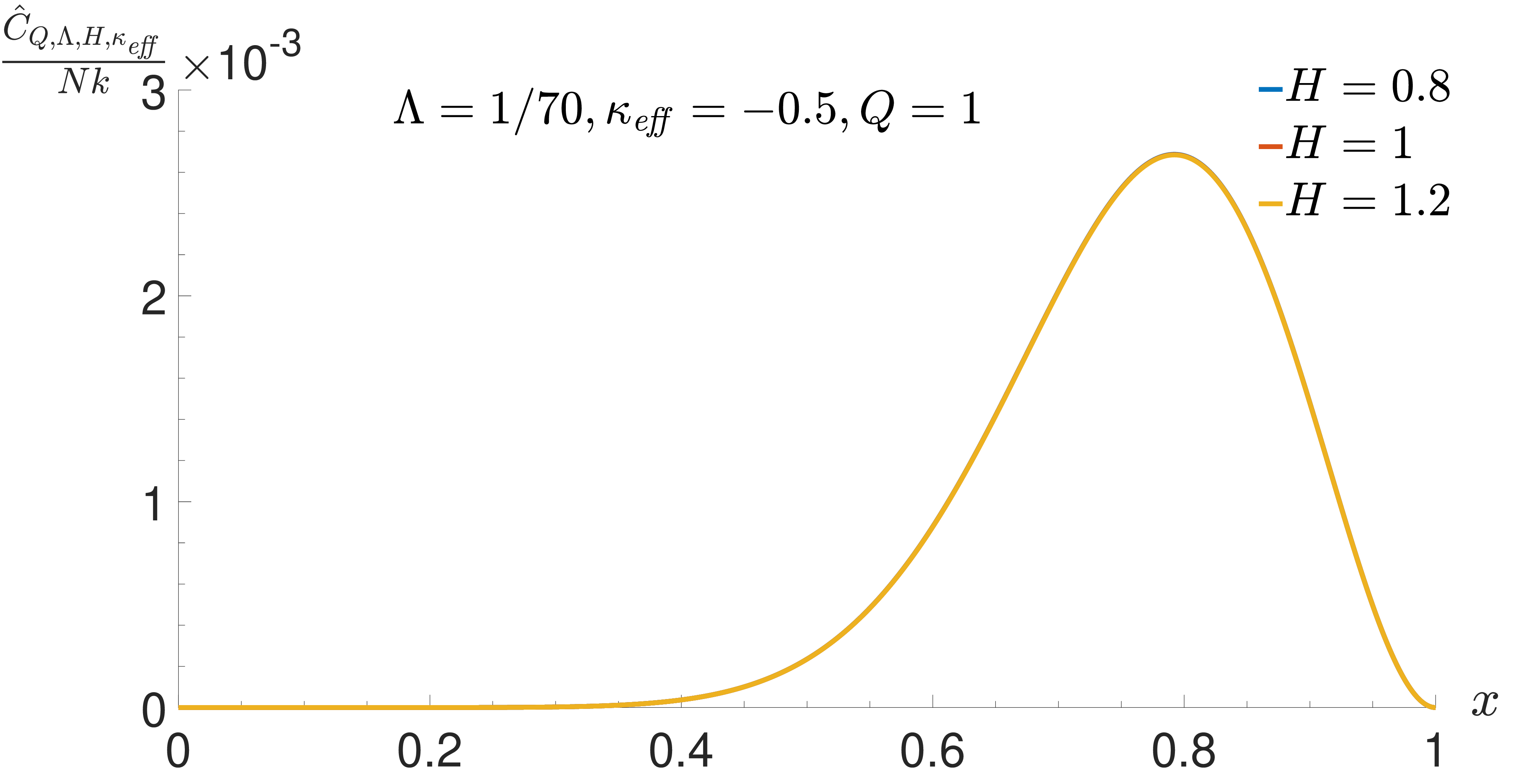}}~
\\
\subfigure[]{\includegraphics[width=8cm,height=4.5cm]{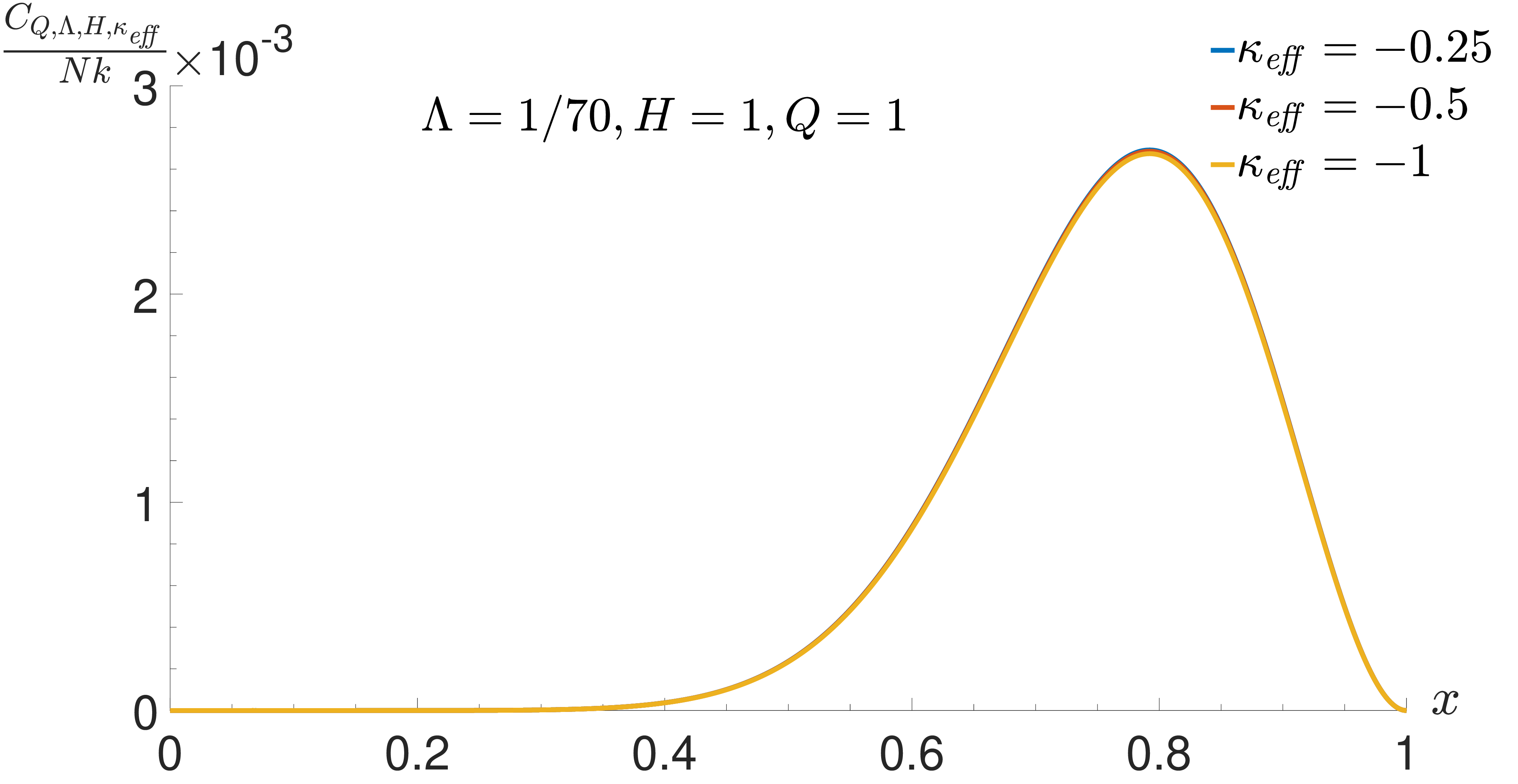}}~
\hspace{0.000001cm}
\subfigure[]{\includegraphics[width=8cm,height=4.5cm]{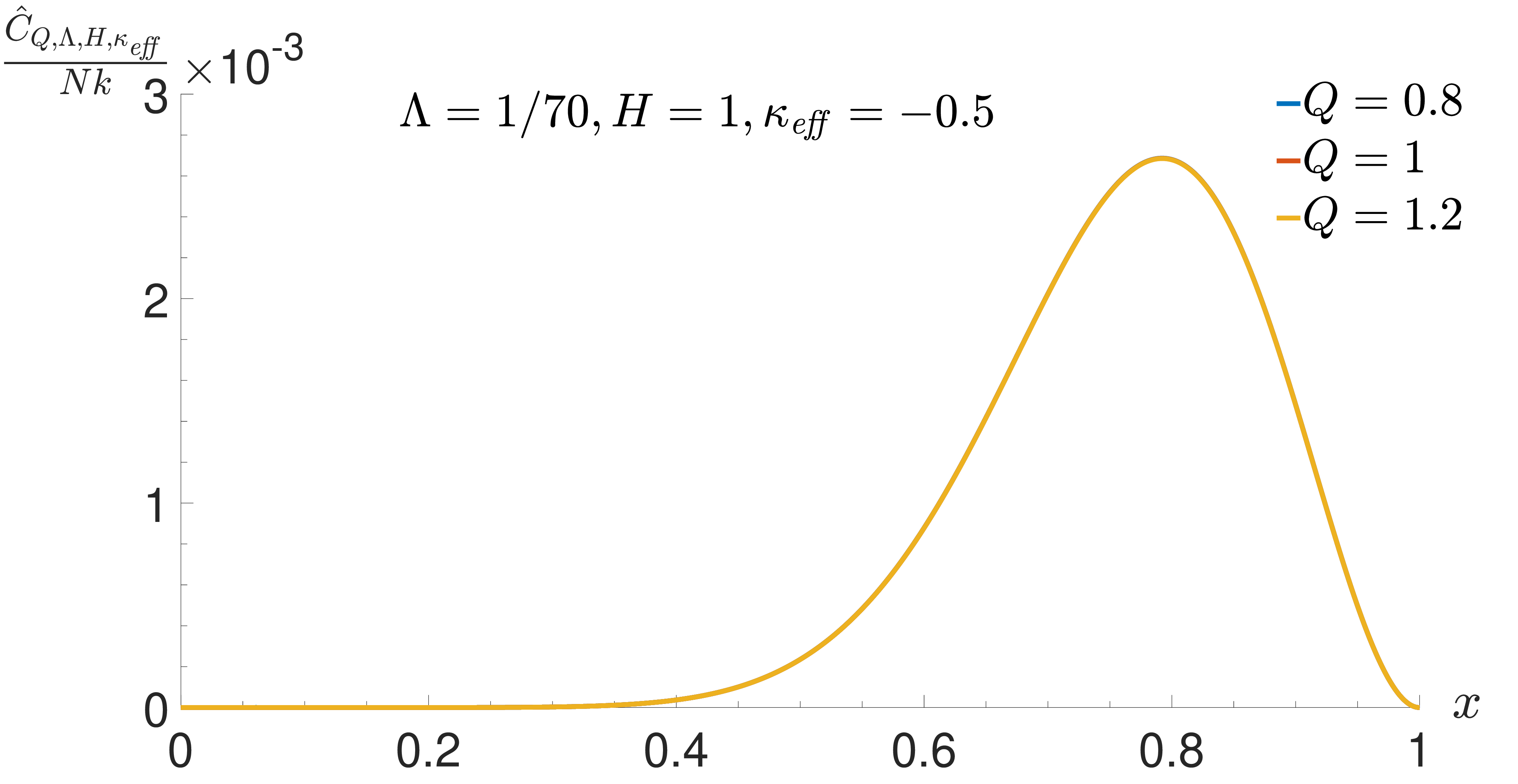}}~
\vskip -1mm \caption{(Color online)The ${{\hat{C}}_{Q,\Lambda ,H,{{\kappa }_{eff}}}}/Nk-x$ curves for different values of the parameters $\Lambda$, $H$, ${{k}_{eff}}$ and $Q$ in the canonical ensemble}
\end{figure}

As shown in Figures 5.2 and 5.3, when the two horizons are regarded as distinct energy levels, the variation of the heat capacity ${{\hat{C}}_{Q,\Lambda ,H,{{\kappa }_{eff}}}}/Nk$ of particles occupying these levels--with either the effective temperature ${{T}_{eff}}$ or the horizon position ratio $x$--follows a trend similar to that of a standard two-level system (Figure 5.1). Specifically, the curve exhibits a maximum, and ${{\hat{C}}_{Q,\Lambda ,H,{{\kappa }_{eff}}}}/Nk\to 0$ both as ${{T}_{eff}}\to 0$ (or $x\to 0$) and as ${{T}_{eff}}\to \infty $ (or $x\to 1$). Moreover, the ${{\hat{C}}_{Q,\Lambda ,H,{{\kappa }_{eff}}}}/Nk-{{T}_{eff}}$ and ${{\hat{C}}_{Q,\Lambda ,H,{{\kappa }_{eff}}}}/Nk-x$ curves are independent of the values of the spacetime state parameters $\Lambda$, $H$, ${{k}_{eff}}$ and $Q$.

As shown in Figures 5.2 and 5.3, the peak values of the ${{\hat{C}}_{Q,\Lambda ,H,{{\kappa }_{eff}}}}/Nk-{{T}_{eff}}$ and ${{\hat{C}}_{Q,\Lambda ,H,{{\kappa }_{eff}}}}/Nk-x$ curves are approximately $2.7\times {{10}^{-3}}$. From Figures 4.1 and 4.2, when $1/\Lambda =70$, the peak values of the ${{C}_{Q,\Lambda ,H,{{\kappa }_{eff}}}}-{{T}_{eff}}$ and ${{C}_{Q,\Lambda ,H,{{\kappa }_{eff}}}}-x$ curves are approximately 4150. The peak values of the curves are independent of the values of $H$, ${{k}_{eff}}$ and $Q$. When $1/\Lambda $ is set to 60 and 80, the corresponding peak values are approximately 3300 and 5100, respectively. Therefore, the relationship between the peak value of ${{C}_{Q,\Lambda ,H,{{\kappa }_{eff}}}}$ and the cosmological constant $\Lambda$ can be approximately expressed as follows:
\begin{align}\label{5.6}
{{C}_{Q,\Lambda ,H,{{\kappa }_{eff}}}}(\emph{peak})\approx \frac{1}{70\Lambda }[4150+31(1/\Lambda -70)]
\end{align}

The ${{C}_{Q,\Lambda ,H,{{\kappa }_{eff}}}}(peak)$ values corresponding to $1/\Lambda $= 60, 70 and 80 were computed using Eq. (5.6), yielding results of approximately 3302, 4150 and 5049 respectively.

A comparison of Eq. (5.6) with the peak values of the ${{\hat{C}}_{Q,\Lambda ,H,{{\kappa }_{eff}}}}/Nk-{{T}_{eff}}$ and ${{\hat{C}}_{Q,\Lambda ,H,{{\kappa }_{eff}}}}/Nk-x$ curves show that:
\begin{align}\label{5.7}
{{\hat{C}}_{Q,\Lambda ,H,{{\kappa }_{eff}}}}\approx \text{2}\text{.7}\times \text{1}{{\text{0}}^{-3}}Nk\sim{{C}_{Q,\Lambda ,H,{{\kappa }_{eff}}}}(peak)\approx \frac{1}{70\Lambda }[4150+31(1/\Lambda -70)]
\end{align}

Thus, for a canonical ensemble, it can be concluded that the number of microscopic particles between the two horizons in their coexistence region is:

\begin{align}\label{5.8}
N\approx \frac{[4150+31(1/\Lambda -70)]}{70\Lambda \times 2.7\times {{10}^{-3}}k}=\frac{[4150+31(1/\Lambda -70)]}{7\Lambda \times 2.7k}\times {{10}^{2}}
\end{align}

Similarly, for the grand canonical ensemble, we plotted the ${{\hat{C}}_{{{\phi }_{eff}},\Lambda ,H,{{\kappa }_{eff}}}}/Nk-{{T}_{eff}}$ and ${{\hat{C}}_{{{\phi }_{eff}},\Lambda ,H,{{\kappa }_{eff}}}}/Nk-x$ curves for different values of the parameters $\Lambda$, $H$, ${{k}_{eff}}$ and ${{\phi }_{eff}}$, as shown in Figures 5.4 and 5.5.

\begin{figure}[htbp]
\centering
\subfigure[]{\includegraphics[width=8cm,height=4.5cm]{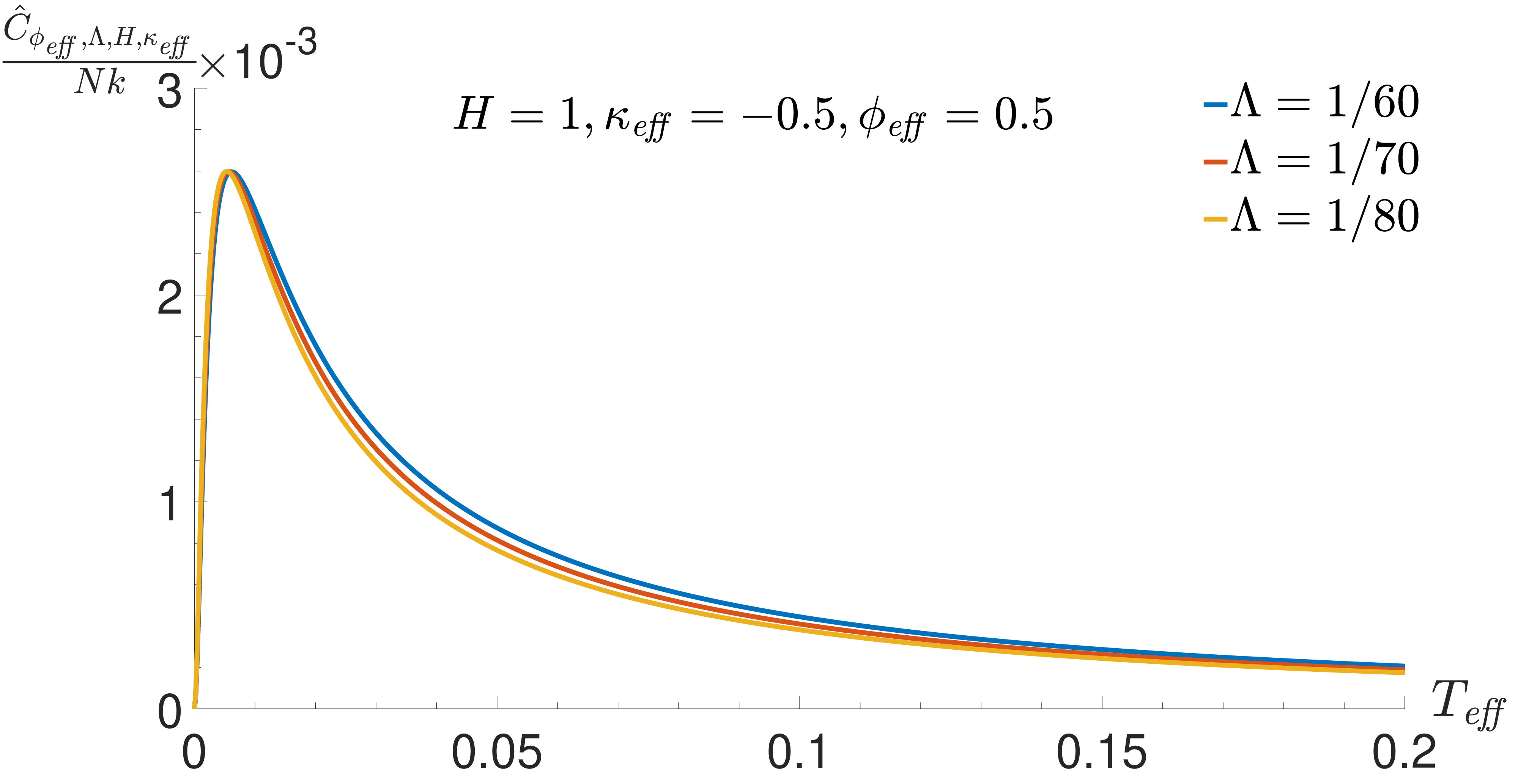}}~
\hspace{0.000001cm}
\subfigure[]{\includegraphics[width=8cm,height=4.5cm]{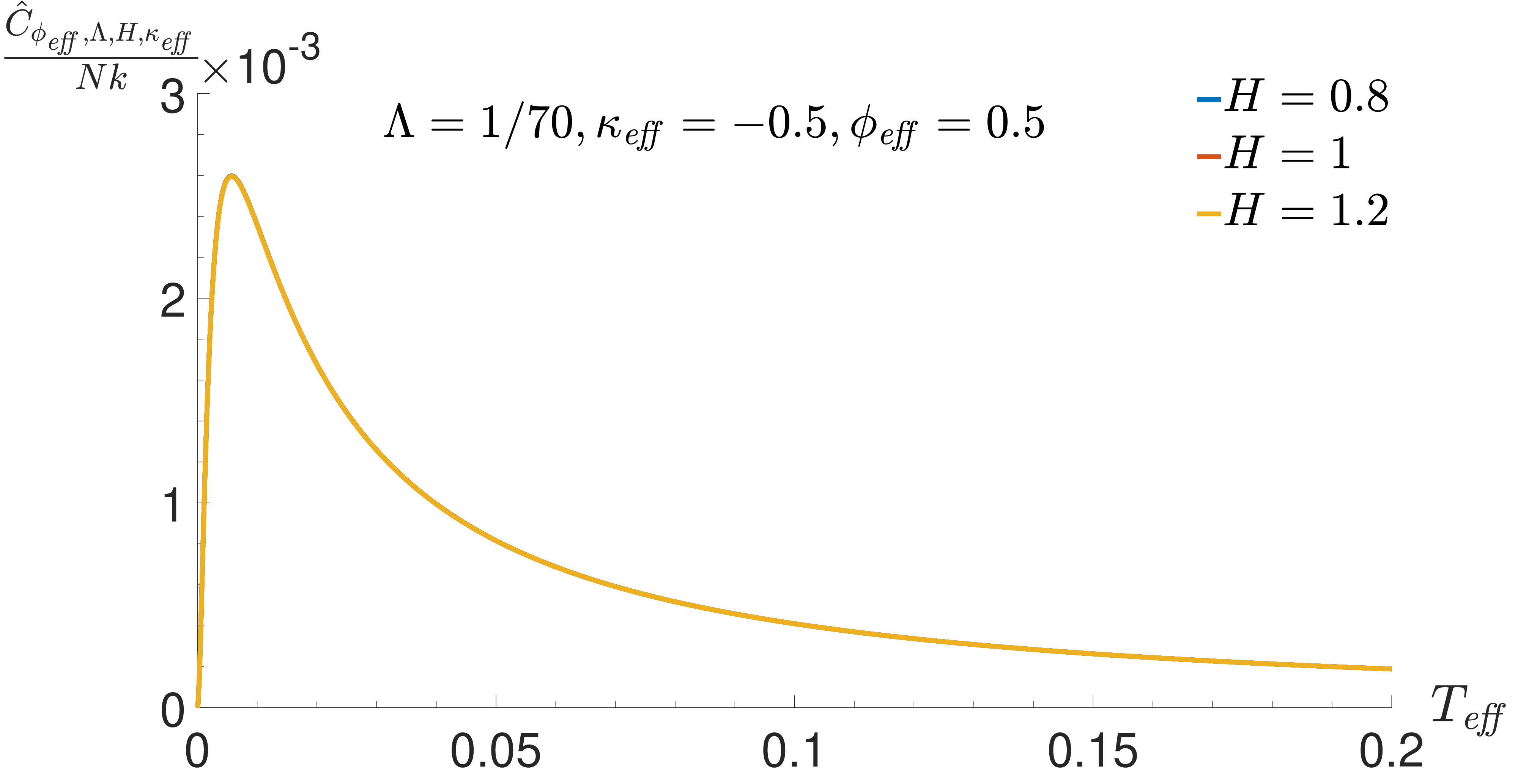}}~
\\
\subfigure[]{\includegraphics[width=8cm,height=4.5cm]{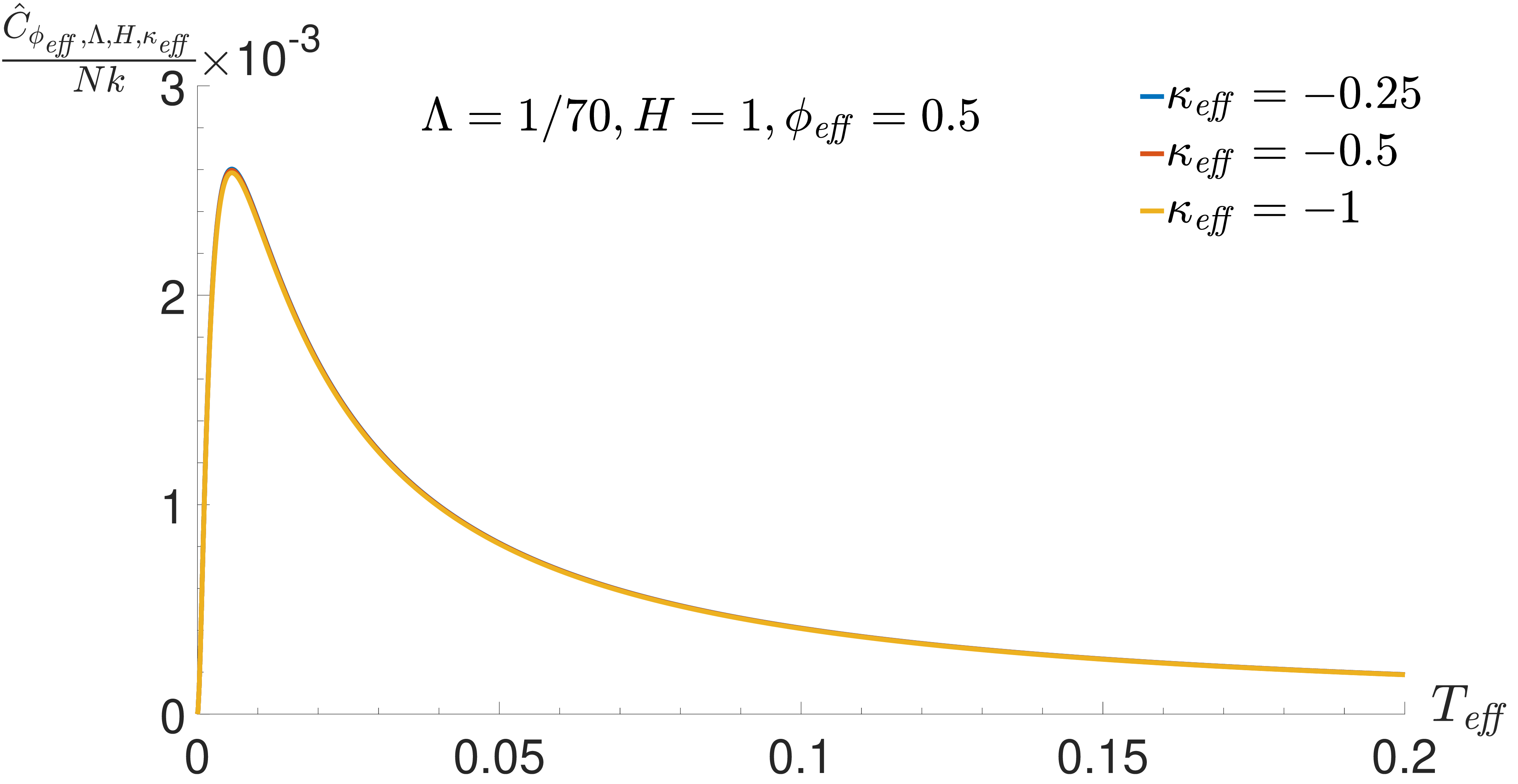}}~
\hspace{0.000001cm}
\subfigure[]{\includegraphics[width=8cm,height=4.5cm]{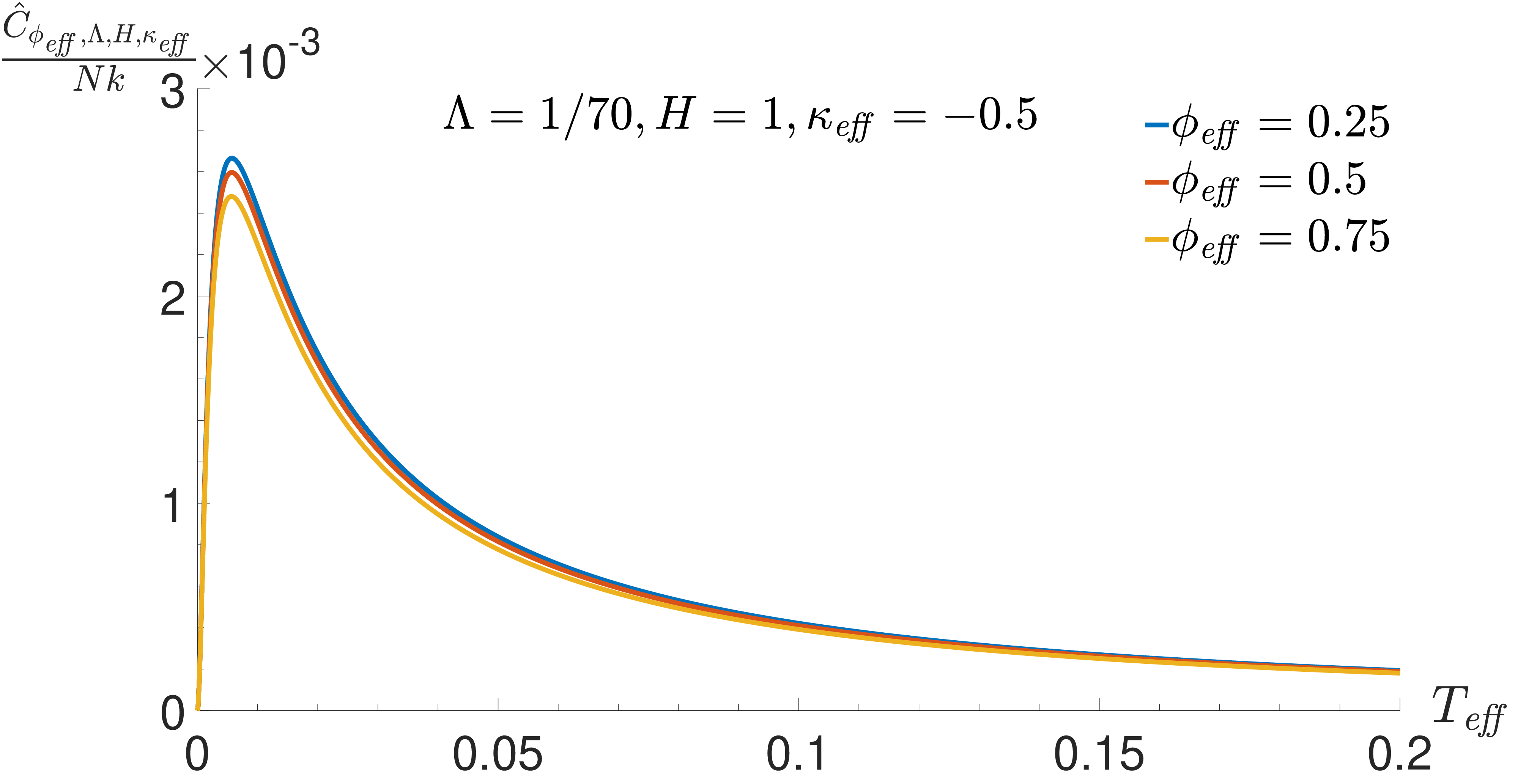}}~
\vskip -1mm \caption{(Color online)The ${{\hat{C}}_{{{\phi }_{eff}},\Lambda ,H,{{\kappa }_{eff}}}}/Nk-{{T}_{eff}}$ curves for different values of the parameters $\Lambda$, $H$, ${{k}_{eff}}$ and ${{\phi }_{eff}}$ in the grand canonical ensemble}
\end{figure}
\begin{figure}[h]
\centering
\subfigure[]{\includegraphics[width=8cm,height=4.5cm]{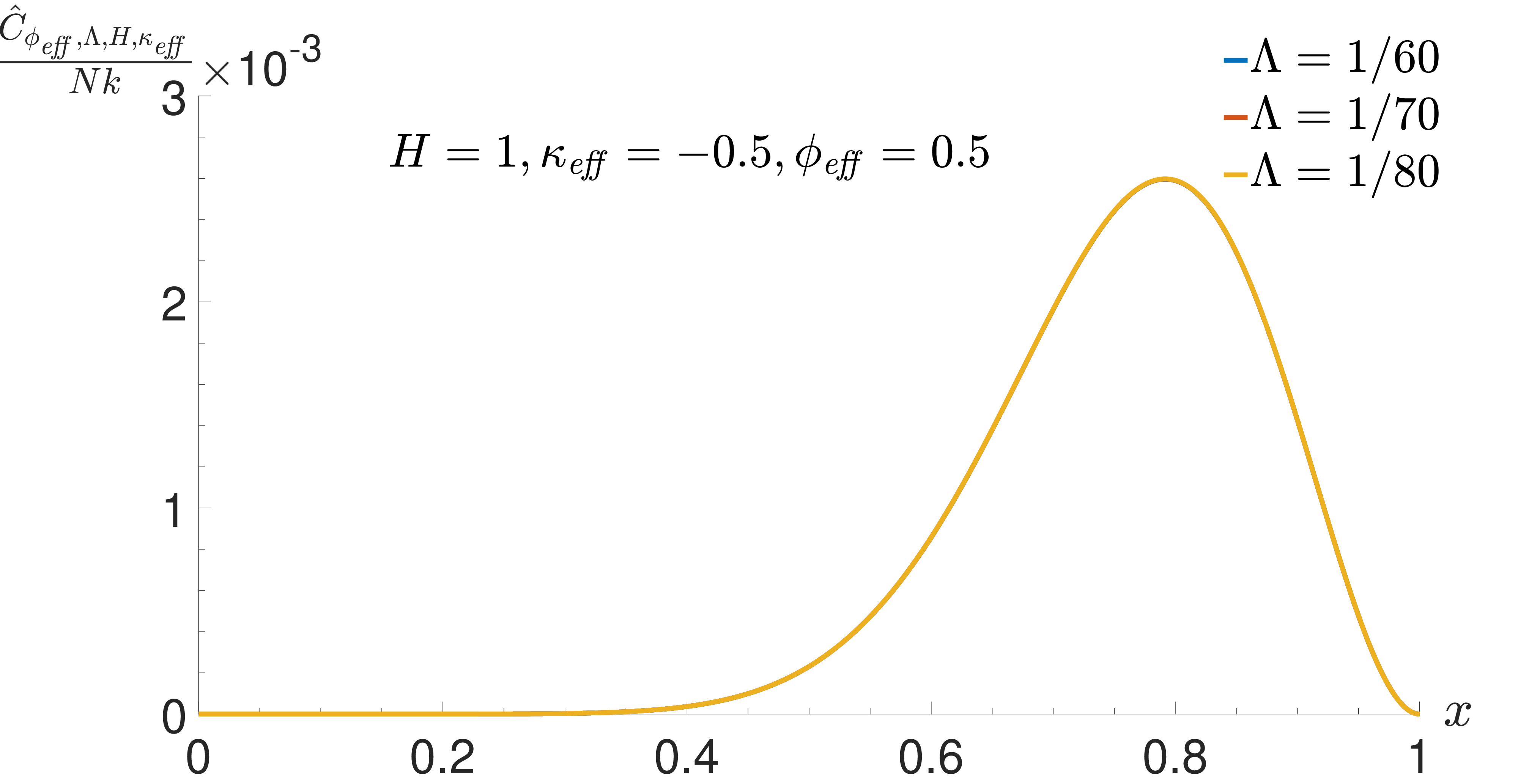}}~
\hspace{0.000001cm}
\subfigure[]{\includegraphics[width=8cm,height=4.5cm]{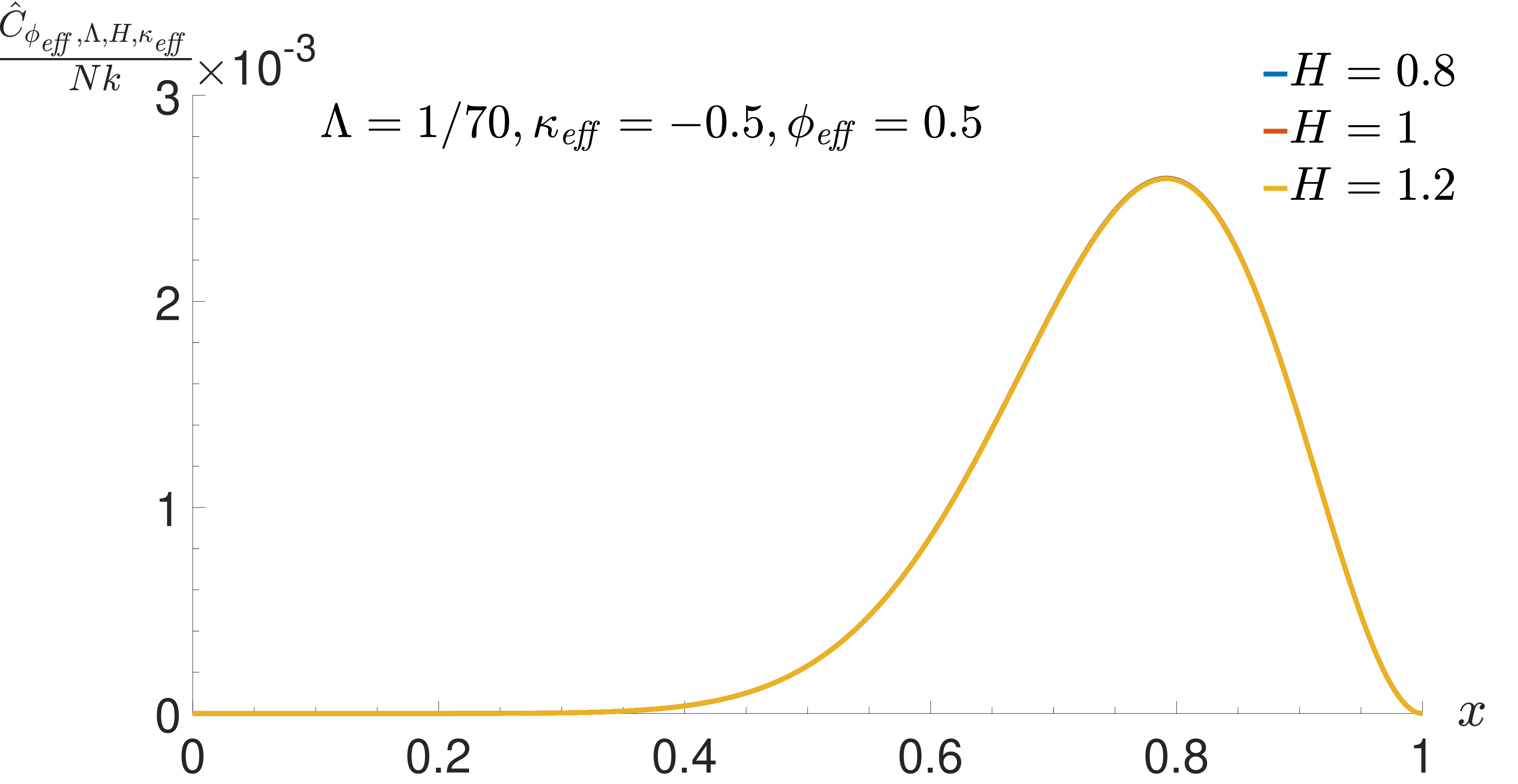}}~
\\
\subfigure[]{\includegraphics[width=8cm,height=4.5cm]{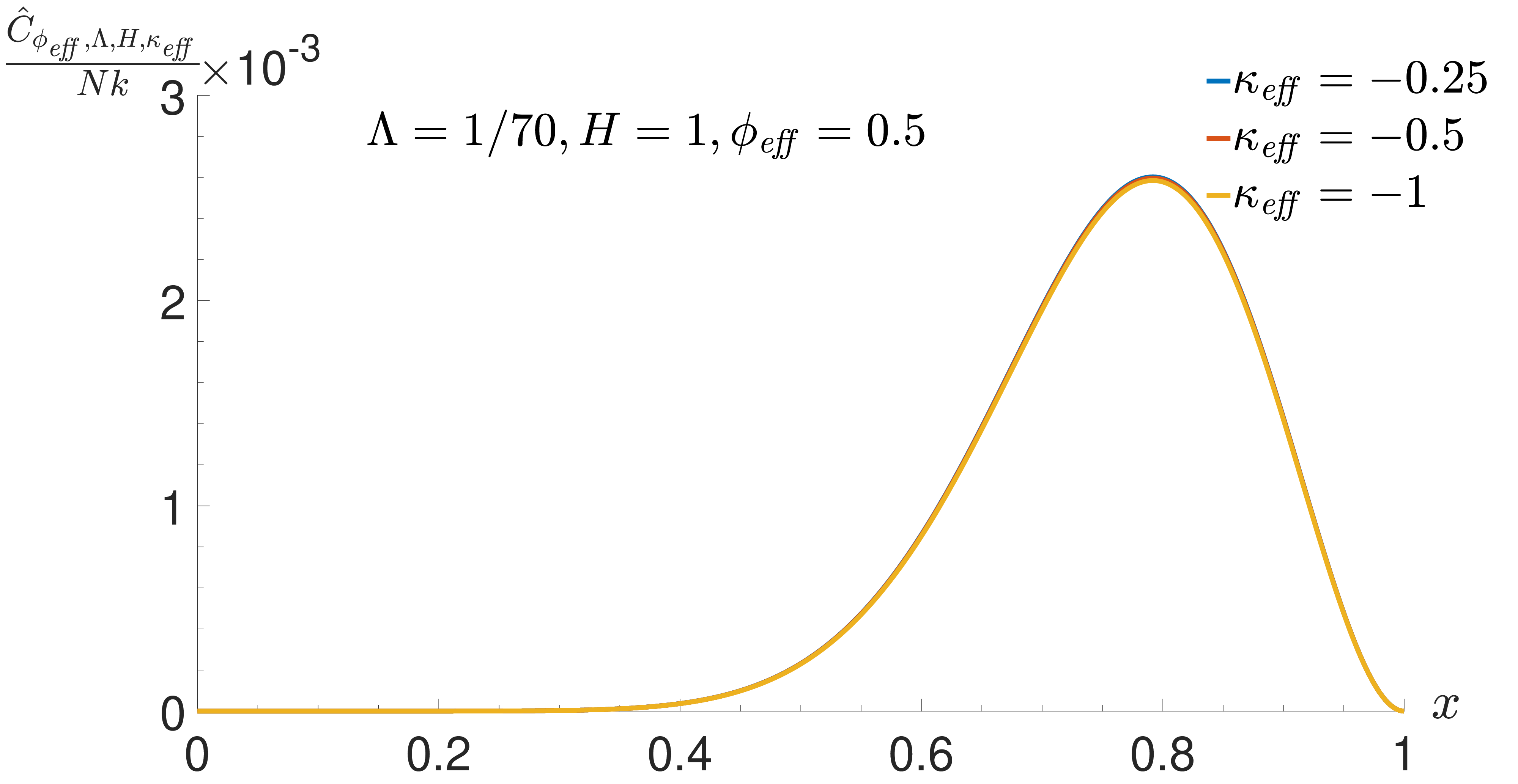}}~
\hspace{0.000001cm}
\subfigure[]{\includegraphics[width=8cm,height=4.5cm]{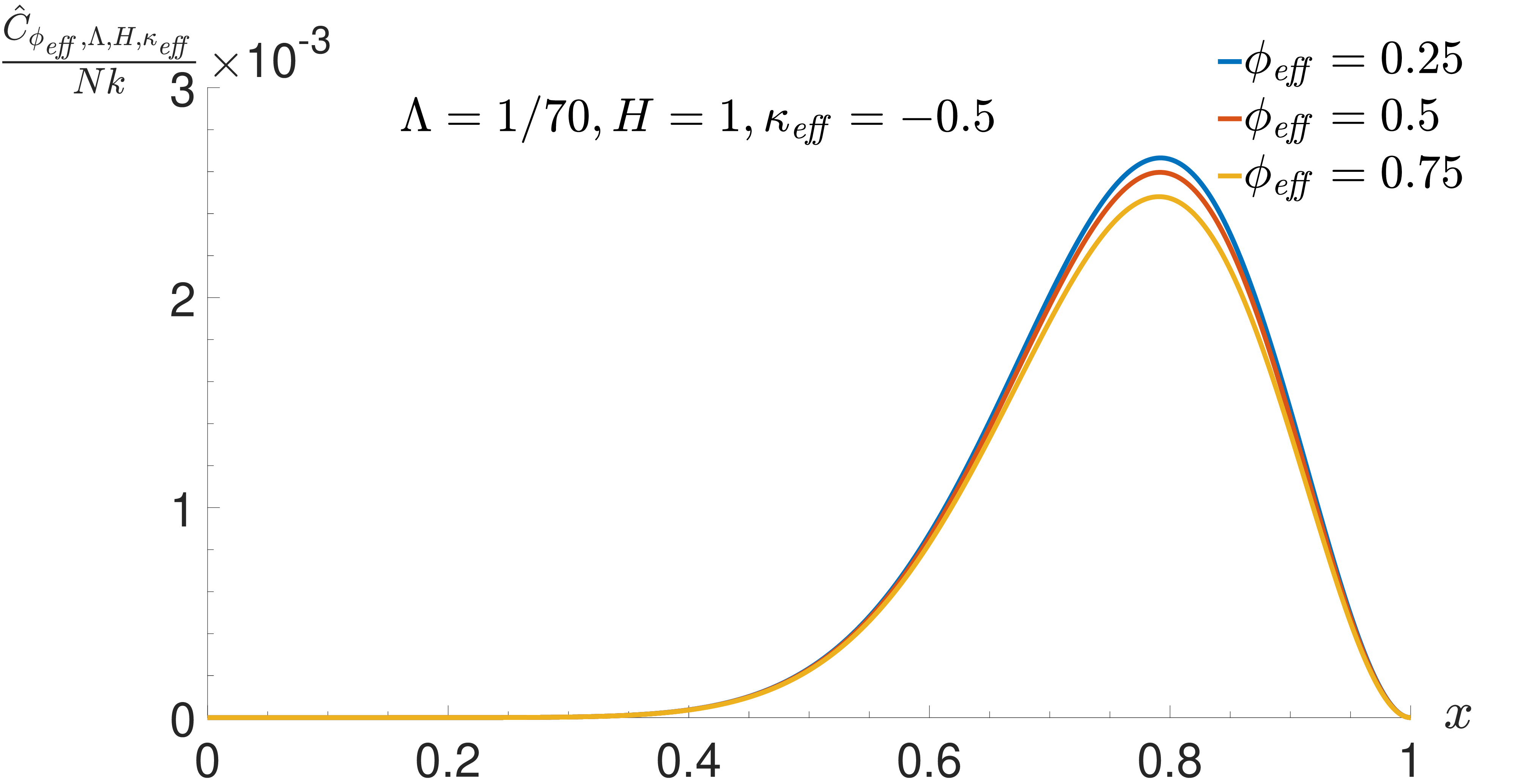}}~
\vskip -1mm \caption{(Color online)The ${{\hat{C}}_{{{\phi }_{eff}},\Lambda ,H,{{\kappa }_{eff}}}}/Nk-x$ curves for different values of the parameters $\Lambda$, $H$, ${{k}_{eff}}$ and ${{\phi }_{eff}}$ in the grand canonical ensemble}
\end{figure}

As shown in Figures 5.4 and 5.5, when the two horizons are regarded as distinct energy levels, the behavior of the heat capacity ${{\hat{C}}_{{{\phi }_{eff}},\Lambda ,H,{{\kappa }_{eff}}}}/Nk$--which varies with the effective temperature ${{T}_{eff}}$ or the ratio $x$ of the horizon positions--along with its maximum value, is independent of the spacetime parameters $\Lambda$, $H$ and ${{k}_{eff}}$. However, the peak value of the curve is influenced by the effective potential ${{\phi }_{eff}}$, as illustrated in Figures 5.4(d) and 5.5(d). Specifically, when ${{\phi }_{eff}}=0.25$, ${{\hat{C}}_{{{\phi }_{eff}},\Lambda ,H,{{\kappa }_{eff}}}}/Nk\sim2.8\times {{10}^{-3}}$; when ${{\phi }_{eff}}=0.5$, ${{\hat{C}}_{{{\phi }_{eff}},\Lambda ,H,{{\kappa }_{eff}}}}/Nk\sim2.7\times {{10}^{-3}}$; and when ${{\phi }_{eff}}=0.75$, ${{\hat{C}}_{{{\phi }_{eff}},\Lambda ,H,{{\kappa }_{eff}}}}/Nk\sim2.6\times {{10}^{-3}}$. That is, under otherwise identical spacetime parameters, the heat capacity ${{\hat{C}}_{{{\phi }_{eff}},\Lambda ,H,{{\kappa }_{eff}}}}/Nk$ decreases near its peak as the effective potential ${{\phi }_{eff}}$ increases, whereas away from the peak region, the value of ${{\hat{C}}_{{{\phi }_{eff}},\Lambda ,H,{{\kappa }_{eff}}}}/Nk$ remains unaffected by ${{\phi }_{eff}}$. This phenomenon, not observed in conventional thermodynamic systems, represents a unique thermodynamic feature of the 5-dimensional de Sitter hairy spacetime. This finding provides new evidence for exploring the distinctions between black hole thermodynamic systems and ordinary thermodynamic systems.

As shown in Figures 4.3 and 4.4, the variation of ${{C}_{{{\phi }_{eff}},\Lambda ,H,{{\kappa }_{eff}}}}$ with the effective temperature ${{T}_{eff}}$ or the horizon position ratio $x$, as well as its maximum value, exhibits no dependence on the spacetime parameters $H$, ${{k}_{eff}}$ and ${{\phi }_{eff}}$, but is related to the cosmological constant $\Lambda$ as follows:

\begin{align}\label{5.9}
{{C}_{{{\phi }_{eff}},\Lambda ,H,{{\kappa }_{eff}}}}(peak)\approx \frac{1}{70\Lambda }[4150+31(1/\Lambda -70)]
\end{align}

The approximate relationship between the peak value of ${{\hat{C}}_{{{\phi }_{eff}},\Lambda ,H,{{\kappa }_{eff}}}}/Nk$ and the effective potential ${{\phi }_{eff}}$ is derived from Figures 5.4(d) and 5.5(d) as:
\begin{align}\label{5.10}
{{\hat{C}}_{{{\phi }_{eff}},\Lambda ,H,{{\kappa }_{eff}}}}/Nk\approx 2.7\times {{10}^{-3}}-\frac{{{\phi }_{eff}}-0.5}{2.5}\times {{10}^{-3}}
\end{align}

Therefore, for a grand canonical ensemble, the number of microscopic particles between the two horizons in the coexistence region, as derived from Eqs. (5.9) and (5.10), can be approximated as:
\begin{align}\label{5.11}
{{N}_{grand}}\approx \frac{1}{7\Lambda k}\frac{[4150+31(1/\Lambda -70)]}{2.7-\frac{{{\phi }_{eff}}-0.5}{2.5}}\times {{10}^{2}}
\end{align}

Conversely, in regions far from the peak of the heat capacity ${{\hat{C}}_{{{\varphi }_{eff}},\Lambda ,H,{{\kappa }_{eff}}}}/Nk$, the approximate number of microscopic particles ${{N}_{grand}}$ between the two horizons in the coexistence region is independent of ${{\phi }_{eff}}$, as seen in Figures 5.4(d) and 5.5(d). This ${{\phi }_{eff}}$-independent behavior underscores a fundamental distinction between black hole microstates and ordinary thermodynamic systems, exposing a unique quantum feature of the 5-dimensional de Sitter hairy spacetime. Understanding the cause of this difference paves the way for deeper investigation into the dynamics of microscopic particles in black holes.

\section{Discussions and Conclusions}\label{six}

This study begins by establishing the effective thermodynamic quantities for the mutually correlated black hole and cosmological horizons in 5-dimensional de Sitter hairy spacetime under specific boundary conditions. These results provide a foundation for further investigating the thermodynamic properties of the coexisting horizon region. Building on this foundation, we analyze the heat capacities of the coexisting horizon region under different conditions. It is found that the ${{C}_{Q,\Lambda ,H,{{\kappa }_{eff}}}}-{{T}_{eff}}$ and ${{C}_{{{\phi }_{eff}},\Lambda ,H,{{\kappa }_{eff}}}}-{{T}_{eff}}$ curves exhibit characteristics reminiscent of Schottky-type specific heat. To uncover the underlying mechanism behind this behavior, we model the black hole and cosmological horizons as two distinct energy levels within an effective thermodynamic system--essentially treating them as a two-level system. The heat capacity of this two-level system is then examined. By comparing Figures 4.1 and 5.2, 4.2 and 5.3, 4.3 and 5.4, as well as 4.4 and 5.5, we observe a remarkable similarity between the heat capacity behavior of the two-level system, that of the effective thermodynamic system, and that of a conventional two-level system (Figure 5.1). This leads to the conclusion that the distinctive features of the ${{C}_{Q,\Lambda ,H,{{\kappa }_{eff}}}}-{{T}_{eff}}$ (${{C}_{{{\phi }_{eff}},\Lambda ,H,{{\kappa }_{eff}}}}-{{T}_{eff}}$) curves in the 5-dimensional de Sitter hairy spacetime are predominantly determined by the heat capacity characteristics of the corresponding two-level system, as described by its ${{\hat{C}}_{Q,\Lambda ,H,{{\kappa }_{eff}}}}-{{T}_{eff}}$(${{\hat{C}}_{{{\phi }_{eff}},\Lambda ,H,{{\kappa }_{eff}}}}-{{T}_{eff}}$) profile.

As shown in Figures 5.2 to 5.5, the quantum system with a split spectrum (i.e., a two-level system) in 5-dimensional de Sitter hairy spacetime exhibits the following behavior: as ${{T}_{eff}}\to 0$, ${{\hat{C}}_{Q,\Lambda ,H,{{\kappa }_{eff}}}}\to 0$ (${{\hat{C}}_{{{\phi }_{eff}},\Lambda ,H,{{\kappa }_{eff}}}}\to 0$), while the specific heat capacity displays a distinct maximum. Moreover, the heat capacity conforms to the universal two-level relation given by Eq. (5.3), which corresponds to the well-known Schottky anomaly. For general two-level systems, their significance lies in the fact that many physical systems can be effectively modeled as two-level systems. In the context of de Sitter spacetime, where there generally exist two horizons with different radiation temperatures, a natural question arises: Does the variation of heat capacity in a system described by effective thermodynamic quantities follow the same behavioral pattern as that of a two-level quantum system-where the two horizons are treated as two energy levels within an effective thermodynamic framework-as observed in the 5-dimensional de Sitter hairy spacetime? In other words, is the heat capacity behavior of the thermodynamic system in 5-dimensional de Sitter hairy spacetime universal across different dS spacetimes?

Furthermore, how do different parameters in various dS spacetimes influence the above behavior? Notably, in the grand canonical ensemble, we observe that near the peak of the heat capacity ${{\hat{C}}_{{{\phi }_{eff}},\Lambda ,H,{{\kappa }_{eff}}}}$, the value of the effective potential ${{\phi }_{eff}}$ has a measurable impact on ${{\hat{C}}_{{{\phi }_{eff}},\Lambda ,H,{{\kappa }_{eff}}}}$, whereas in other regions, ${{\hat{C}}_{{{\phi }_{eff}},\Lambda ,H,{{\kappa }_{eff}}}}$ remains largely unaffected by variations in ${{\phi }_{eff}}$. What is the underlying physical reason for this phenomenon? A deeper investigation into these questions will contribute to a more comprehensive understanding of the quantum properties of black holes. Moreover, it opens new research directions for exploring the thermodynamic characteristics of dS spacetimes.

\section*{Acknowledgments}
We would like to thank Prof. Ren Zhao for his indispensable discussions and comments. This work was supported by the Natural Science Foundation of Shanxi Province (202303021211180) and the Program of State Key Laboratory of Quantum Optics and Quantum Optics Devices (KF202403).

\end{document}